# Climatic extremes in Brazil: a parallel analysis of historical trends and socioeconomical impacts

## Extremos climáticos no Brasil: uma análise em paralelo de tendências históricas e impactos socioeconômicos

**Davi Lazzari[I], Amália Buchweitz Garcez[I], Nicole Magalhães Poltozi[II], Gianluca de Souza Pozzi[II], Carolina Brito[I]**

[I] Universidade Federal do Rio Grande do Sul, Porto Alegre, RS, Brasil
[II] Universidade do Vale do Rio dos Sinos, São Leopoldo, RS, Brasil

## ABSTRACT

An important consequence of human-induced climate change is the increase in extreme weather events. This study contributes to the understanding of Brazil's climate change by examining historical temperature and precipitation patterns. Extreme events of temperature and precipitation are identified using data from the Brazilian Institute of Meteorology, which includes records from 634 meteorological stations operating intermittently since 1961. Using the first 30 years (1961–1990) as the reference period, our results show a significant increase in warm days and a corresponding decrease in cold days over the last 30 years (1991–2020), in agreement with previous works. In terms of precipitation, it indicates a trend toward drier conditions in the Northeast region of Brazil, whereas the South is experiencing wetter conditions, with an increase in the number of heavy precipitation days in South and in the extremely dry periods in the Northeast. These results have been verified for consistency with several extreme climate indices measured in this study. Additionally, data from S2iD is analyzed, an official database that records natural disasters in Brazil, to estimate their impact in terms of human losses and financial costs over the past decade. Our findings indicate that drought events are the most economically costly, with multiple instances causing damages exceeding a billion USD, whereas storms have the greatest impact on people. Although it is not possible to directly attribute the natural disasters recorded in the S2iD database to the extreme weather events identified through meteorological data, discussion is done on potential implications of these events in the frequency and location of the disasters.

## RESUMO

Uma importante consequência das mudanças climáticas induzidas pelo homem é o aumento dos eventos climáticos extremos. Este estudo contribui para a compreensão das mudanças climáticas no Brasil ao examinar padrões históricos de temperatura e precipitação. Eventos extremos de temperatura e precipitação são identificados utilizando dados do Instituto Nacional de Meteorologia do Brasil, que incluem registros de 634 estações meteorológicas operando intermitentemente desde 1961. Usando os primeiros 30 anos (1961–1990) como período de referência, os resultados mostram um aumento significativo de dias quentes e uma diminuição correspondente de dias frios nos últimos 30 anos (1991–2020), em concordância com trabalhos anteriores. Em termos de precipitação, observa-se uma tendência de condições mais secas na região Nordeste do Brasil, enquanto o Sul está experimentando condições mais úmidas, com um aumento no número de dias de precipitação intensa no Sul e nos períodos extremamente secos no Nordeste. Esses resultados foram verificados quanto à consistência com vários índices climáticos extremos medidos neste estudo. Além disso, foram analisados dados do S2iD, uma base de dados oficial que registra desastres naturais no Brasil, para estimar seu impacto em termos de perdas humanas e custos financeiros na última década. Os achados indicam que os eventos de seca são os mais custosos economicamente, com vários casos causando danos superiores a um bilhão de dólares, enquanto as tempestades têm o maior impacto sobre as pessoas. Embora não seja possível atribuir diretamente os desastres naturais registrados na base de dados do S2iD aos eventos climáticos extremos identificados por meio de dados meteorológicos, as possíveis implicações desses eventos na frequência e localização dos desastres são discutidas.

## 1 INTRODUCTION

The effects of human-induced climate change are already being felt in various parts of the world through increasing extreme events, which are large deviations of a climatic state. Global warming has notably impacted the frequency and intensity of severe precipitation and, in certain regions, has led to agricultural and ecological droughts (Masson-Delmotte et al., 2021). Furthermore, projections indicate that if global warming reaches the 2ºC mark compared to the current average temperature, extreme temperature events such as heatwaves and cold waves, which used to occur about once a decade, may become four times more frequent, while extreme events that occurred once every fifty years may become nine times more frequent. These projections more than double in a heating scenario of 4ºC (Masson-Delmotte et al., 2021).

The impacts of extreme events have been extensively studied across different regions of the globe, highlighting the growing risks associated with climate change (Sippel et al., 2015; Otto, 2017; Ebi et al., 2021). For instance, economic losses of USD 2 billion were reported due to extreme rainfall in Beijing in 2012 (Zhang et al., 2013), and severe precipitation in Pakistan in 2010 resulted in over 1,800 fatalities (Solberg, 2010). The 2021 Emergency Events Database (EM-DAT) report attributes over 10,000 deaths, 101.8 million people affected, and approximately USD 252 billion in economic losses worldwide to extreme events in 2021 (CRED, 2022). In Rio Grande do Sul, a state in the South of Brazil, extreme rainfall between late April and early May 2024 affected over 2 million people, representing 20% of the state's population. The catastrophe caused nearly 200 deaths, left approximately 1,000 people injured (Governo do Estado do Rio Grande do Sul, 2024; Reboita et al., 2024) and the financial losses were reported to be in the billions of dollars (World Meteorological Organization, 2024; Caleffi et al., 2024).

Several studies in South America (Haylock et al., 2006; de los Milagros Skansi et al., 2013; Regoto et al., 2021; LagosZúñiga et al., 2024) and in specific regions of Brazil (Dufek e Ambrizzi, 2008; Silva Dias et al., 2013) have used meteorological stations to investigate climate extremes, consistently identifying an increase in extreme temperatures (Rosso et al., 2015; Almeida et al., 2017; Costa et al., 2020) and extreme precipitation events (Murara et al., 2019; Zilli et al., 2016; Ávila et al., 2016; Dufek e Ambrizzi, 2008; de Medeiros et al., 2022) in recent years in Brazil. These observational patterns are further contextualized by regional climate modeling efforts, which highlight large spatial and inter-model variability in precipitation extremes across South America and stronger agreement for warming trends in temperature extremes (LagosZúñiga et al., 2024). More comprehensive analysis of temperature and precipitation extremes were conducted over Brazil by Avila-Diaz et al. (2020) using high-resolution climate datasets, and Regoto et al. (2021) using meteorological stations. The first analysis compares four different datasets, including observational data and reanalysis products, applying multiple indices to analyze extreme events in the period from 1980 to 2016, while the second uses the INMET database from 1961 to 2018 and analyses extreme indices, similarly to what will be done on this work. The results corroborate the consistent warming trends found, characterized by an increase in warm extremes and a decline in cold extremes. As for precipitation, the findings show an increase in consecutive dry days and a reduction in consecutive wet days across most regions of Brazil.

The present work aims to review and integrate previously reported trends in climatological extremes observed in Brazil (e.g., Regoto et al., 2021; Alvarez-Diaz et al., 2020) and to quantitatively estimate the human and economic losses in the last 12 years. Specifically, trends in extreme air temperature and rainfall events are analyzed using data from the Brazilian National Institute of Meteorology (INMET, 2022), covering over 60 years of observations from conventional weather stations. In parallel, the impacts of officially recorded natural disasters are explored, as compiled by the Integrated Disaster Information System (S2iD, UFSC, 2025).

Unlike previous studies that focus on either meteorological trends (e.g., Zilli et al., 2016; Costa et al., 2020) or disaster impacts (e.g., Minervino and Duarte, 2016), the contribution of the present work lies in the simultaneous examination of these two aspects to characterize the broader landscape of climate-related risks in Brazil. Moreover, some differences in respect to previous studies in terms of methodology and data presentation are introduced. A broad and complementary set of precipitation indices are employed —including R20mm, DD90, SPI, SPEI, and mRAI—which enables the evaluation of both the intensity and duration of dry and wet conditions. Also, trends are detected using an index that compares the average frequency of extreme events between two 30-year periods: 1961–1990 (a reference period) and 1991–2020. Although this work does not establish a causal relationship between extreme events and disasters, presenting both datasets side by side allows to illustrate how patterns of climate extremes and disaster impacts have evolved concurrently, offering context for future studies on climate adaptation and risk management.

This work is structured as follows: In Section 2, a detailed discussion of the data utilized and the methods employed in the analysis is presented. Section 3 focuses on the results, encompassing both the meteorological station data and the analysis of natural disasters. Both the sections of methodology and results are complemented by the data in the Supplementary Material (SM). Finally, the conclusions of the study are drawn in Section 4.

## 2 METHODS

Brazil is the fifth largest country in the world in terms of area, with more than 200 million people according to the last census (IBGE, 2022). It covers 8,514,876 km and possesses an important diversity of climates, as shown by the Köppen-Geiger Climate clasification presented in Figure 1. Brazil is geopolitically divided into five regions, shown in the inset image on the left, in Fig. 1. Although the regional division used in this study is based primarily on political and administrative boundaries rather than strict climatic criteria, it still holds meteorological relevance, and is especially meaningful for analyzing socioeconomic conditions.

### 2.1 INMET Database

The Brazilian Institute of Meteorology (INMET, 2022), is the government agency responsible for collecting and producing reports on meteorological data. To identify the average trends and extreme events (EE) of temperature and precipitation, this study used data from the INMET meteorological stations shown as white circles in Fig. 1. This database is composed of 634 conventional meteorological stations in all Brazilian regions, with various operating periods and data gaps. The stations are mostly located in regions with a higher population density, resulting in a non-homogeneous distribution across the territory.

Most of the stations started collecting data on January 1st, 1961, and have been doing so intermittently throughout the subsequent 60-year period. The data was considered up to December 31st, 2020. From this database the following variables were used: daily maximum air temperature (TX), daily minimum air temperature (TN) and daily rainfall (also referred as precipitation, PR) time series. The 60 years of data were split into a reference period (RP: 1961–1990) and an analysis period (AP: 1991–2020) to assess pattern changes.

Figure 1 – Map of Brazil

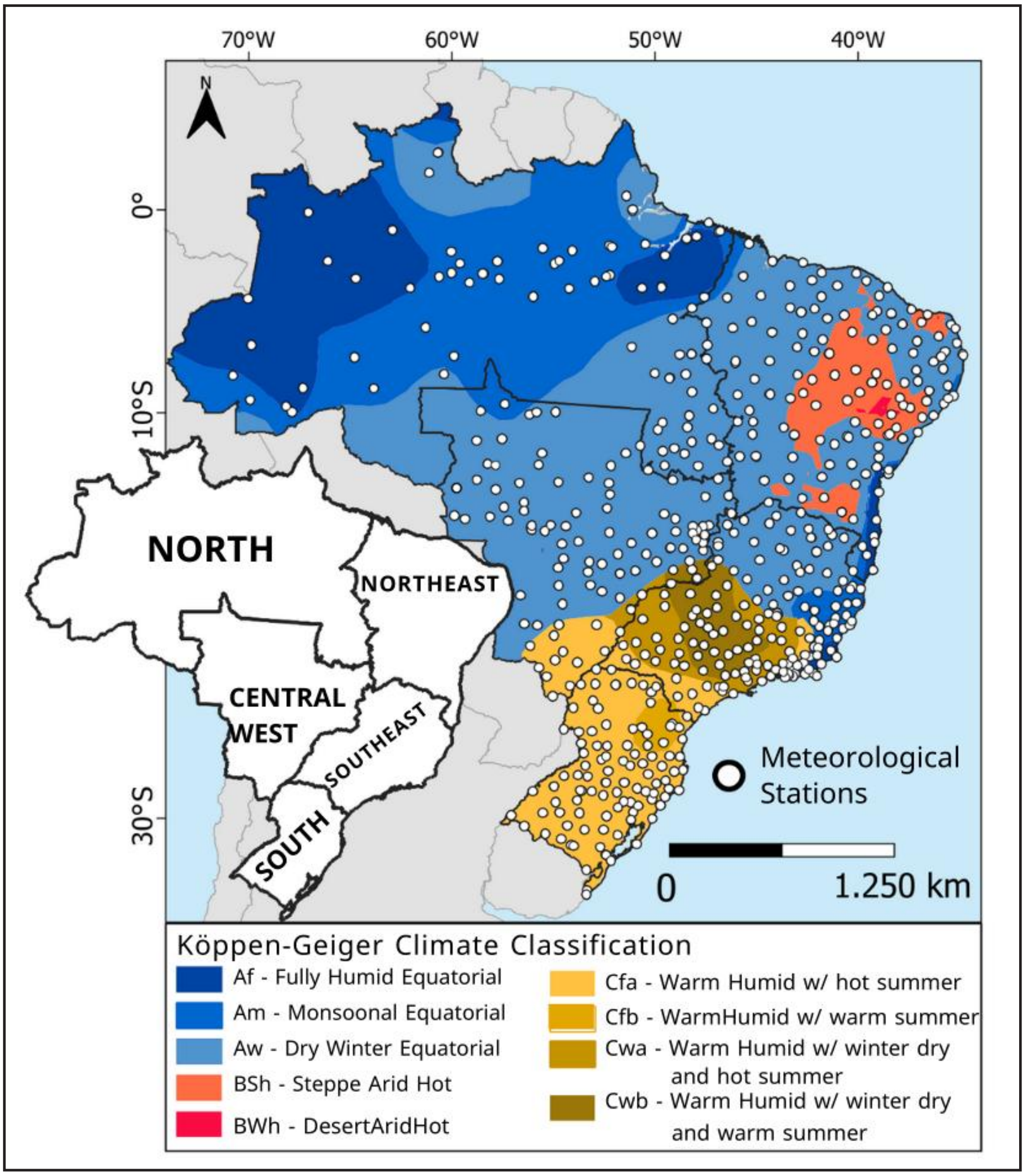

Caption: Map of Brazil with different colors indicating the Köppen-Geiger Climate classification, and white points the INMET meteorological stations. The inset on the left shows the five official Brazilian geopolitical regions. Source: Authors' collection (2025)

### 2.1.1 Data Treatment

In the analysis, stations with less than 30% of data during the AP and the RP were disregarded, and the years where an individual station worked less than 30% of the year were removed, which resulted in 287 working stations. This fraction threshold was defined after several tests with functioning cuts from 10-50%, considering that different cuts could imply changes in the EE identification. In the SM, the tests for the historical series of the indices TX90p, R20mm and DD90 (that will be presented in the next section) are shown, as well as a comparison between the 10% and 50% threshold cuts in the R20mm index for all stations (in a map distribution). The test showed a mostly robust behavior between the thresholds. The 30% cut was chosen in order to maintain a balance between the amount of data and the robustness of our results, as it retains the general trends of different threshold curves. Additional cleanings on the data series are done, such as exclusion of outliers, defined as values that are significantly more extreme than typical observations for each station based on the spread of the local data. Formally, an outlier is identified when it exceeds the 75th percentile by more than three times the interquartile range. Two normalization parameters are also defined to account for missing data and determine the yearly mean EE over all stations (described in the next section and in detail in the SM).

### 2.1.2 Extreme climate indices

To investigate the air temperature behavior, the TN and the TX month anomalies were computed using average yearly values. To retrieve the temperature extremes, a series of known indices proposed by the Expert Team on Climate Change Detection and Indices (Zhang et al. 2004; Zhang et al. 2011; https://etccdi.pacificclimate.org/list_27_indices.shtml) were regarded. The set of indices chosen for the temperature analysis are detailed in Table 1.

For the precipitation and drought analyses in Brazil, the Standard Precipitation Index (SPI; McKee et al., 1993; Svoboda et al., 2012) and the Standard Precipitation

Evapotranspiration Index (SPEI; Vicente-Serrano et al., 2010) were used at time scales from 3 to 12 months. The SPEI differs from the SPI in that it incorporates temperature to estimate Potential Evapotranspiration (PET), which is then subtracted from precipitation to calculate a water balance. Both indices were calculated using the Climpact2 software (https://climpact-sci.org/). In addition, several well-known climate indices were computed (Zhang et al., 2005, 2011; Tank et al., 2009; Rooy, 1965; Haensel et al., 2015; Regoto et al., 2021), and an Extremely Dry Period (DD90) was defined as a sequence of days without rain exceeding the 90th percentile for each station (Table 2). The Rainfall Anomaly Index (RAI), introduced by van Rooy (1965), measures how wet or dry a given period (usually a month or year) is compared to long-term historical rainfall records. The mRAI index was chosen instead of the RAI for comparison with SPI and SPEI because it uses a scaling factor m, as proposed by Haensel et al. (2015). The scaling factor adjusts the magnitude of RAI values so they are on a comparable numerical scale with other drought/wetness indices such as SPI and SPEI. Table 2 lists the indices used to analyze rainfall and drought, along with their definitions and where it is presented, either in the main text or in the SM.

Table 1 – Temperature Indices

| Index | Index Name | Definition | Unit | Paper location |
|---|---|---|---|---|
| TN10p | Cold Nights | Days when TN < 10th percentile. | days | Main |
| TN90p | Warm Nights | Days when TN > 90th percentile. | days | SM |
| TX10p | Cold Days | Days when TX < 10th percentile. | days | SM |
| TX90p | Warm Days | Days when TX > 90th percentile. | days | Main |
| WSDI | Warm Spell Duration Indicator | Events with ≥ 6 consecutive days with TX > 90th percentile. | days | SM |
| DTR | Diurnal Temperature Range | Annual mean difference between daily TX and TN. | ºC | SM |

Source: Authors' private collection (2025). Acronym, name, definition, unit and location in the paper. TN refers to the daily minimum temperature, TX represents the daily maximum temperature and the numbers 10 and 90 in the first four indices refer to the chosen percentile threshold of the data. The last two indices' acronyms are simply the index name initials

To calculate the relative changes between extremes the RP (1961-1990) and AP (1991-2020), Equation 1, the anomaly of extreme events (ΔEE), was adapted from Bador et al. (2018). This index is useful in understanding the geographical distribution of EE, and is defined for each meteorological station,

$$\Delta EE_s = \left(\frac{\overline{EE}_{AP,s}}{\overline{EE}_{RP,s}} - 1\right) \times 100\%, \tag{1}$$

Where $\overline{EE}_{AP,s}$ is the mean number of EE in each meteorological station *s* averaged over the 30 years in the AP while $\overline{EE}_{RP,s}$ is averaged over the 30 years in the RP. Notice that this fraction is greater than 1 if the number of EE increases in the AP compared to the RP, and less than 1 otherwise. To centralize this value around zero, 1 is subtracted from it and the result is multiplied by 100% to obtain the EE percentage increase or decrease in the AP compared to the RP. This index is used for individual stations for the temperature and precipitation extremes. All the percentile thresholds were calculated only with data from the RP.

To take into account the fact that each meteorological station does not always operate for an entire year, the number of EE detected is divided by the fraction of functioning days in each year. After summing over all the yearly EE in a given period, the AP, this number is divided by the quantity of years the station worked in the same period. This number is divided by the mean number of EE in a year, calculated over the RP in the same manner as the AP. These treatments are explained mathematically in the SM.

Table 2 – Rainfall Indices

| Index | Index Name | Definition | Unit | Paper location |
|---|---|---|---|---|
| DD90 | Extremely Dry Period | Series of consecutive dry days ≥ 90th percentile of the dry days distribution. | days | Main |
| RAI | Rain Anomaly Index | Difference between yearly precipitation in a given year and mean yearly precipitation in the RP divided by the difference between the mean of the 10 most or least rainy years and the mean yearly precipitation for positive or negative anomalies, respectively. | unitless | SM |
| mRAI | Modified Rain Anomaly Index | Same as the RAI index with a different scaling factor (1.7) (Cardona et al., 2012) | unitless | Main |
| SPI | Standard Precipitation Index | Measure of how current precipitation compares to historical averages by converting rainfall data into a standardized scale, it defines whether conditions are wetter or drier than normal. | unitless | Main |
| SPEI | Standard Precipitation Evapo-transpiration Index | Measure of how current precipitation and potential evapotranspiration compare to historical averages by converting the water balance into a standardized scale, it defines whether conditions are wetter or drier than normal. | unitless | Main |
| R20mm | Heavy Precipitation Days | Days where daily precipitation ≥ 20 mm. | days | Main |
| R50mm | Very Heavy Precipitation Days | Days where daily precipitation ≥ 50 mm. | days | SM |
| R100mm | Extremely Heavy Precipitation Days | Days where daily precipitation ≥ 100 mm. | days | SM |
| SDII | Simple Precipitation Index | Total precipitation divided by the number of wet days (daily rainfall ≥ 1 mm) in a year. | mm/day | SM |
| PRCPTOT | Annual Precipitation | Total precipitation in a year (normalized). | mm | SM |

Source: Authors' private collection (2025). Acronym, name, definition, unit and location in the paper. In the first index, DD refers to Dry Days – that is, a distribution of days without rain – and the following number refers to the data percentile threshold (90%). The four following indices' acronyms (RAI, mRAI, SPI and SPEI) are the index name initials. The next indices refer to rainfall of over XX mm, RXXmm (XX = 20, 50 or 100) and PRCP in the last index refers to precipitation

## 2.2 S2iD Database

The information on natural disasters was obtained from the Brazilian Integrated System of Disasters (S2iD) (UFSC, 2025). The database assembles data on cities affected by disasters, compiling the activities following the events, their damages and risks. This is useful to follow the situations in individual cities, as well as for any needed declarations or acknowledgments of an emergency or a state of public calamity. While the S2iD is an official database for natural disasters in Brazil and has been widely used in studies (Dalagnol et al., 2022; Ramos Filho, 2021; Minervino e Duarte, 2016), it has several limitations - as will be commented below - and cannot be considered a comprehensive resource for analyzing the temporal and spatial dynamics of these events across the country.

The public policy on natural disasters in Brazil was consolidated as a civil defense system in 2005 (Kuhn et al., 2022). Currently, however, the only systematic organization of information on disasters comes from the S2iD platform, from 2012 to the present date, where each disaster is recorded manually by municipal and state authorities and then recognized by the federal government. Notably, the S2iD database presents some challenges such as data incompleteness, duplicity of events, records not individualized by municipalities and by type of disaster (Kuhn et al., 2022; Carvalho, 2018). To mitigate these issues, events reported in the same city on the same day were taken as a single event (considering always the most costly, human or economic-wise), reducing the dataset from 65,079 to 41,462 events. As such, the possibilities that the database offers after the cleaning outweigh its fault, enabling to access information with a reasonable degree of reliability regarding human and economic losses from disasters.

The S2iD contains information on more than 17 types of disasters. In this study, the focus was on the natural kinds of disasters, which were additionally grouped into 4 categories: storm, flood, drought and ”others”, the later including several disasters that, despite generally being related to extremes, are less frequent in the S2iD, such as landslides, forest fires (less common until the year 2020), heat and cold waves, dam failures, among others, as shown in Figure 1 in the SM. For each disaster, there are more than 40 damage parameters to be filled in a form, such as number of deaths, number of injured, number of dislodged, cost of public material damage and cost of private losses (agriculture, livestock and other losses).

In 2020, the “Diseases” disaster type, driven by the COVID-19 pandemic, greatly increased in terms of human impact and economic losses. As this type of disaster is better evaluated through the Brazilian Health Ministry’s website (https://covid.saude.gov.br) and is unrelated to extreme climate events, it was excluded from this analysis.

## 3 RESULTS

### 3.1 Air Temperature

One well established possible consequence of human-induced climate change is the increase in the frequency and intensity of extreme weather events. In Figure 2, it is shown that this increase is already happening in Brazil. Black lines in Figs.

2-a and 2-b show the anomaly in TN and TX respectively, in respect to the RP (indicated in the figure), averaged over all meteorological stations. The red lines are a smoothed version of the black line and serve to guide the eyes (calculated using a standard local polynomial regression fit). This figure shows only two of the extreme climate indices detailed in Table 1, as it highlights the changes in extremes in Warm days and Cold nights. The other indices, such as Cold days and Warm nights, presented similar trends and are therefore discussed in the SM.

Figure 2 – Temperature anomalies and extreme events

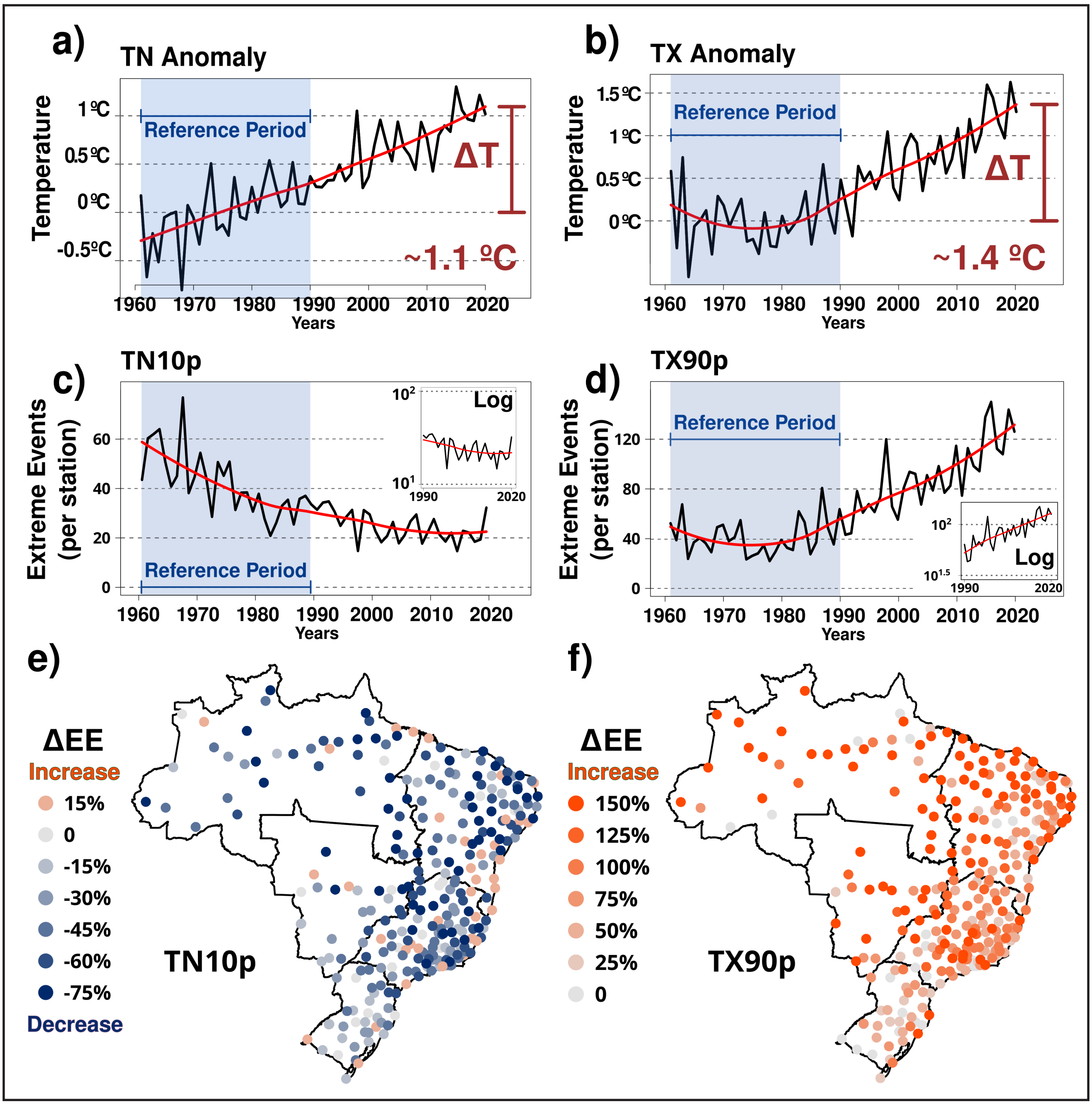

Source: Authors' collection (2025). Caption: a) and b) respectively present black lines depicting the minimum (TN) and maximum (TX) air temperature anomalies (mean temperature deviations). The TN anomaly reaches around 1.1C in the year 2020, while the TX anomaly reaches around 1.4. c) and d) depict the mean yearly extreme events occurrence as TN10p time series, shown to be decreasing, and TX90p, increasing over the whole period of analyses. Both figures present insets with the same data in log-scale and overall trends. These four-time series are averaged over all meteorological stations, and red lines are a smoothed version of the black curves (calculated through standard local polynomial regressions). Panels e) and f) show respectively the TN10p (cold nights) ΔEE and TX90p (warm days) ΔEE (calculated through Eq. 1), which respectively present decreasing (dark blue points) and increasing (orange points) behaviors throughout the country in the analysis period (AP, 1991 to 2020), when compared to the reference period (RP, 1961 to 1990)

The behavior indicates an increase in temperature anomalies, a trend that is also observed at a global scale (Rohde e Hausfather, 2020) and has already been reported in the Amazon (Almeida et al.,2017) and Northeast regions (Costa et al., 2020). The maximum air temperature anomaly in Brazil reaches around 1.4ºC and the minimum temperature anomaly reaches around 1ºC. In the SM, the TN and TX anomalies per month averaged over five years are also shown, indicating a consistent increase from 1960 and 2020 by month, with more pronounced anomalies during the months from June to August.

In Figures 2-c and 2-d, extreme temperature events TN10p and TX90p (also called Cold Nights and Warm Days respectively) are shown as a time series averaged over all meteorological stations. The inset graphs represent the data in the AP in a logarithmic scale that present the same overall behavior as the non-logarithmic graphs. The red lines show that TN10p events are decaying almost linearly in time, while TX90p events are increasing at a slightly higher rate. The increase in TN and TX temperature anomalies as well as the decrease of TN10p and the increase of TX90p indicate a shift of the whole daily temperature distribution to higher temperatures. This trend can also be seen in Figures 2-e and 2-f, which show the spatial distribution of the temperature EE anomaly defined in Eq. 1. This index shows a decrease in TN10p and a significant increase of TX90p, reproduced throughout all Brazilian regions, except for the South where the trend seems to be milder.

### 3.2 Rainfall

In this study, the daily precipitation data from the INMET database (INMET, 2022) was used to calculate the SPI (McKee et al., 1993) and SPEI (Vicente-Serrano et al., 2010) indices through the Climpact software (https://climpact-sci.org/) and further understand the precipitation behavior in Brazil (details of these indices are on SM). The overall SPEI and SPI time series trends were calculated using the Mann-Kendall (MK) test (Mann, 1945) to validate any monotonic trend on the series with a significance level < 5%, and when affirmative, the Sen's slope (Sen, 1968) algorithm is used to obtain the rate of change. Additionally, the mRAI index was also calculated using MK test and Sen's slope to obtain the change rates. In this analysis, positive and negative values represent trends towards wetter and drier conditions, respectively, in the last 60 years. In Fig. 3-a, these indices are shown averaged for each Brazilian region. The upper part of this figure shows the political borders of each region, while the lower part displays the respective average trends for the SPI and SPEI indices measured on time scales ranging from 3 to 12 months, along with the average trend of the mRAI index measured on an annual scale (the full map for each index can be found in the SM).

It is interesting to note in Fig. 3-a that all regions show a robust trend, independently of the time scale in which they are measured and for all the indices. Because the averages are taken over geopolitical regions and some higher trend values can be smoothed out. Nevertheless, the indices show a clear sign of increase in drought in the Northeast region and in precipitation in the South region. The Central-West and Southeast regions also present a trend towards drier weather while the North region shows a milder tendency towards an increased precipitation. The drought in the Northeast and Central-West and the precipitation trends in the South align with Chagas et al., 2022, who identified significant streamflow changes in Brazil's water cycles using data from 886 hydrometric stations and a different methodology. The trend in precipitation for the South region measured in the present work is also in agreement with Schossler et al., 2018, where the authors monitor the precipitation pattern using the Tropical Rainfall Measuring Mission (TRMM) in the period from 1998 to 2013.

Figure 3 – Trends and extreme precipitation events

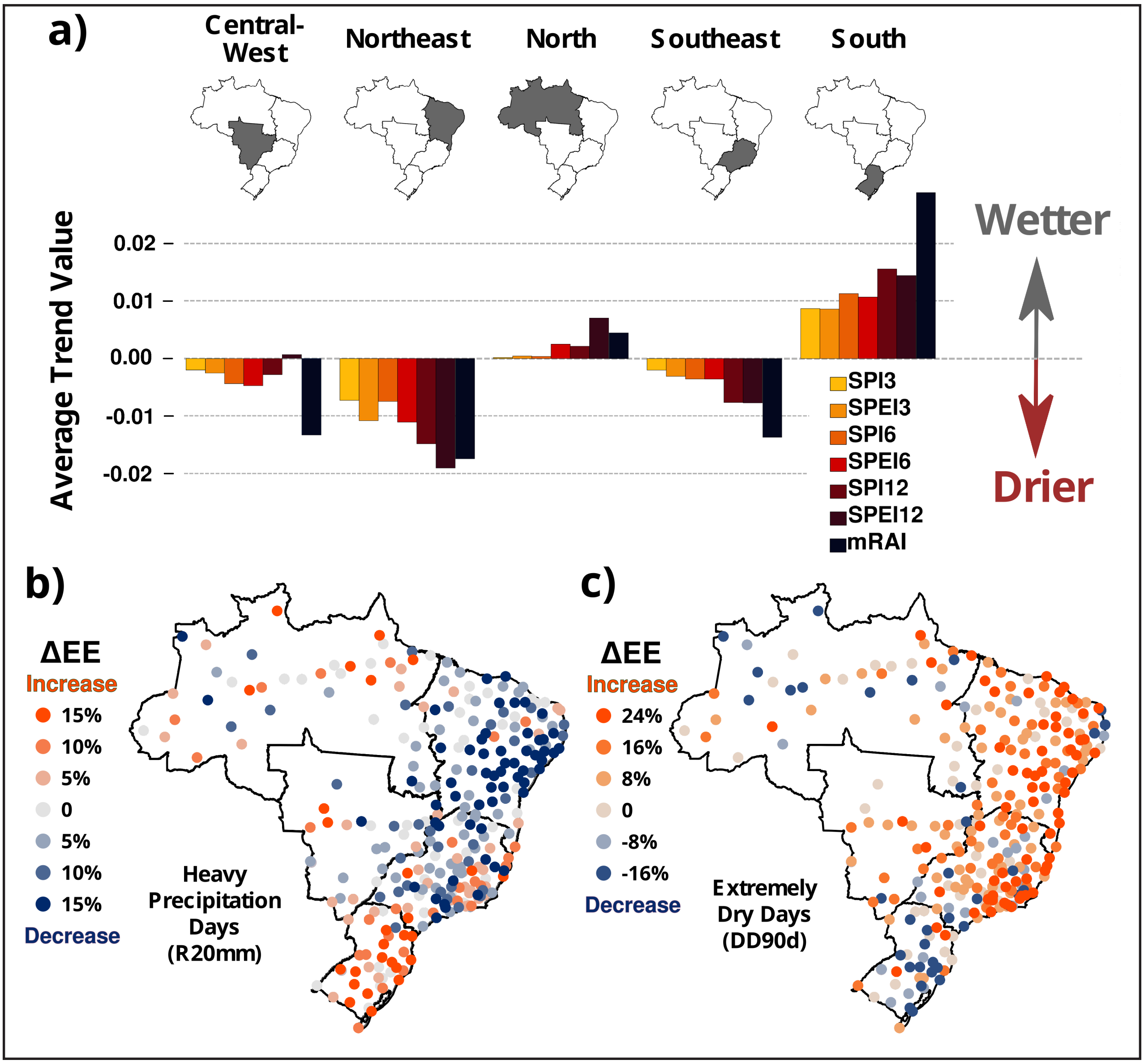

Source: Authors' collection (2025). Caption: a) The SPI, SPEI and mRAI average trends on humidity for each region, with time spans of 3, 6 and 12 months for first two indices. b) Anomaly of Heavy Precipitation Days (R20mm) for each meteorological station, which seems to decrease in the Northeast region while increasing in the South region and c) anomaly of Extreme Dry Period events, shown to be decreasing in the South region and increasing in the Northeast and Southeast. In both figures the EE anomaly, ΔEE, is calculated through equation 1 for individual meteorological stations. The orange color indicates an increase and darkblue a decrease in ΔEE

Figure 3-b shows the spatial distribution of the anomaly of R20mm, ΔEE, defined in Eq. 1. The map with circles representing individual meteorological stations shows an increase in R20mm in the South and a milder decrease in events in the Northeast. The Southeast region between the South and Northeast acts as a transition zone where it is possible to notice mixed behaviors in stations. This overall behavior is in accordance with the results obtained by Avila-Diaz et al., 2020 and Regoto et al., 2021. In particular, studies in the state of Rio de Janeiro—characterized by complex topography and dense urbanization—have identified increasing trends in extreme daily rainfall and decreasing trends in dry/wet spell persistence, contributing to greater irregularity in rainfall distribution and heightened hydrometeorological risk (LuizSilva & OscarJúnior, 2022). Figure 3-c presents the spatial distribution of the anomaly of DD90 EE, also defined through Eq. 1, showing a clear increase in events in the Northeast and Southeast regions, and a decrease in the South region. As done with the temperature analysis, only two of the precipitation indices detailed in Table 2 are shown here. All the additional indices showed similar trends and are detailed in the SM.

The next section will discuss a parallel database that presents human and economic costs of natural disasters. While it does not directly link the EE trends with disasters, it gives an overview of the potential costs that such events can have by looking for the disasters that happened over the last decade in Brazil.

### 3.3 Economic Losses and Human Impact of Natural disasters

This subsection reports the results from the S2iD database (UFSC, 2025), used to evaluate the impacts of disasters in Brazil over the past decade.

The bars in Fig. 4-a show the total number of individuals affected (including deaths, injuries, dislodgement, missing people and others) by type of disaster from 2013 to 2024. Fig. 4-b presents the total economic cost (accounting for all types of financial costs, including material damage, agriculture, livestock, private and public damage) by type of disaster. Figure 5 shows the geographical distribution of natural disasters that have the most significant human impact and cause the largest economic losses, along with EE identified over the same period.

Figure 4 – Human and economic impacts of disasters

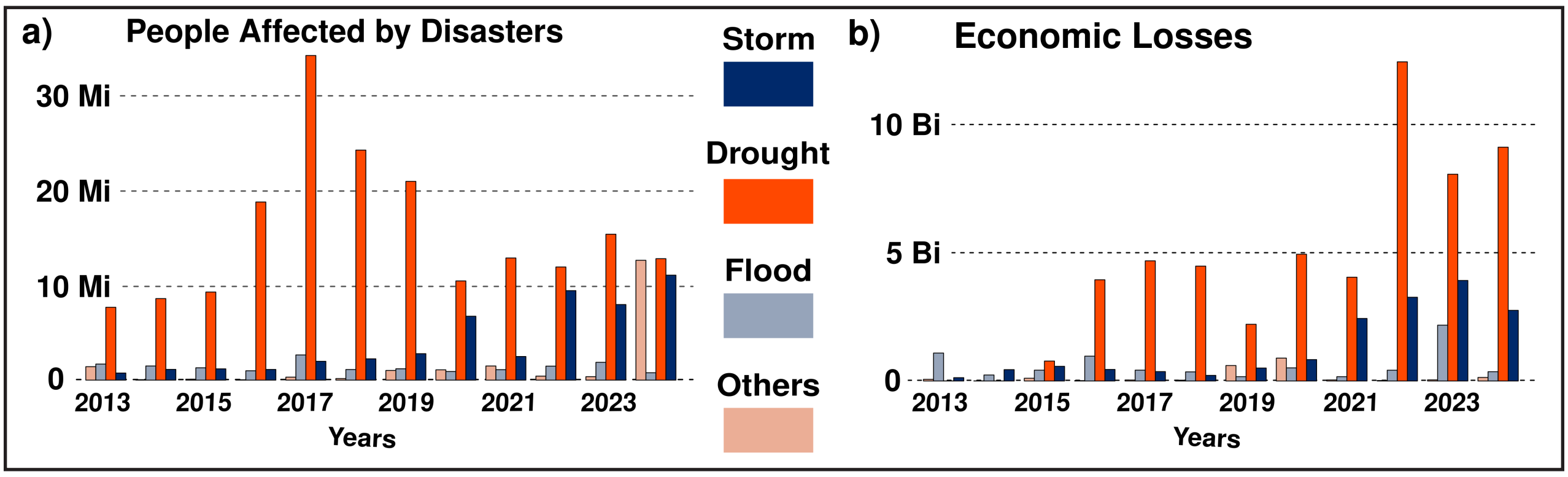

Source: Authors' collection (2025). Caption: a) Number of people affected per year from 2013 to 2024, by type of disaster. b) Economic losses in USD billions per year by type of disaster

The total cost of disasters in Brazil during the analyzed period is approximately USD 80 billion, which represents about 5% of Brazil's annual GDP (around USD 1.8 trillion in 2023 (IBGE, 2023)). In terms of human impact, approximately 270 million people were affected—more than the country's entire population. This suggests that some individuals were impacted multiple times by the disasters, which aligns with the record of several events affecting certain locations in the S2iD database. For instance, the location with the highest number of disasters recorded almost 200 events during the analysis period. The frequency of data entries for the 50 most affected locations is shown in the SM. In addition to being the type of event that affects the most people, drought also stands out as the costliest catastrophic event registered from 2013 to 2023, accounting for approximately USD 66 billion during the period analyzed.

Figures 5a–c shows the geographical distribution of these damages across Brazil, while Figure 5d presents the distribution of EE identified over the same period. Figure 5-a highlights drought as the most impactful event in terms of economic costs across nearly all regions of Brazil, followed by storms, which have a significant effect in the Southeast, South and Central West regions. Figure 5-b shows the geographic distribution of events affecting more than 100,000 people, with drought again being the most impactful in the Northwest region and storms in the South and Southeast. Notably, 'others' (as cited above) are the most devastating disasters

in terms of people affected in the Central-West and North regions, often related to fires. Figure 5-c depicts a specific type of social impact, the internal displacement, clearly showing that storms and floods are the most impactful disasters in forcing people to leave their homes.

Figure 5 – Geographical location of disasters

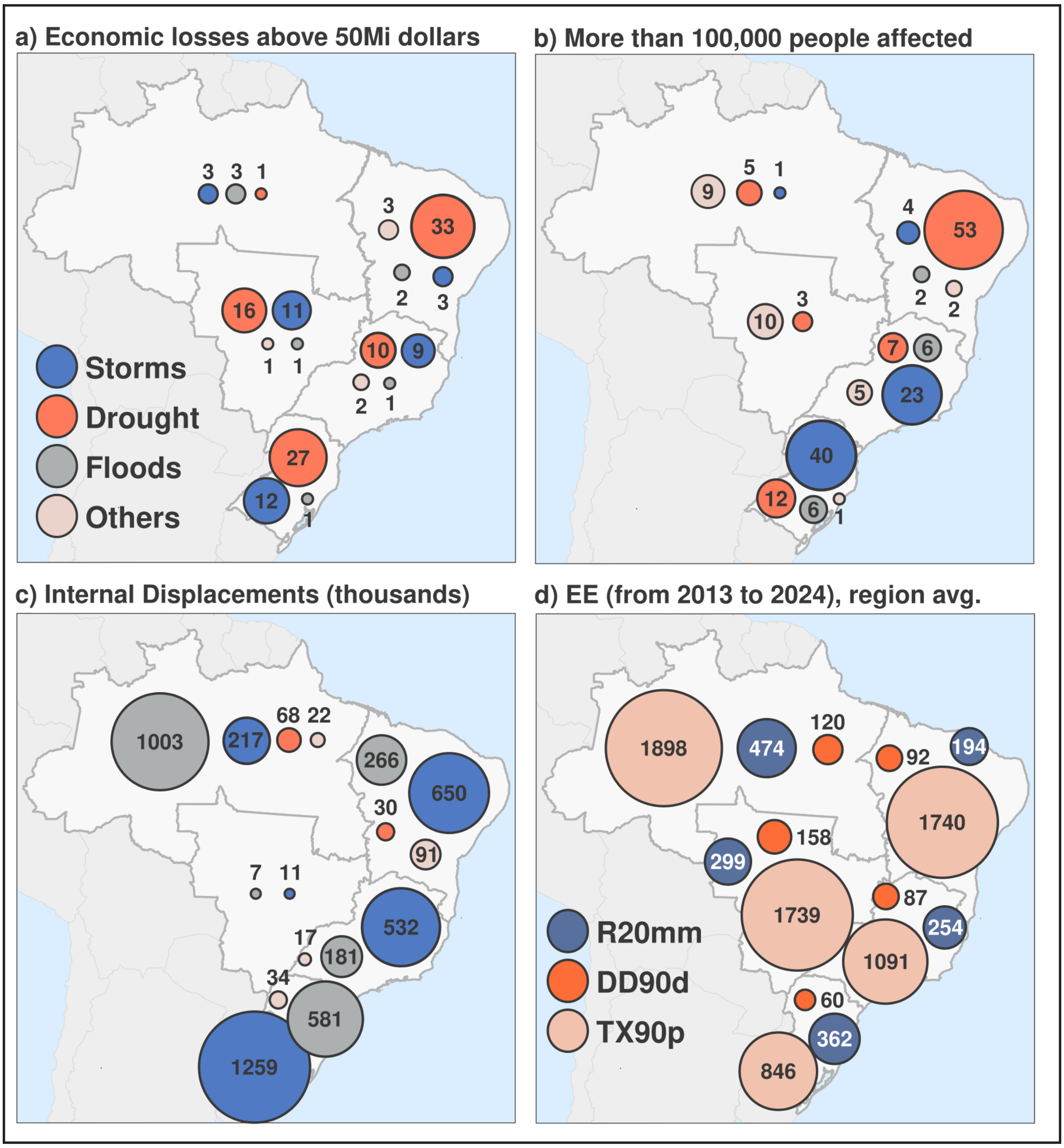

Source: Authors' collection (2025). Caption: a) Number of disasters causing over 50 million USD in damages, b) Number of disasters affecting more than 100,000 people, c) Internal displacements in thousands of people and d) average number of extreme events by meteorological station (from INMET). The discs represent the total number of events recorded from 2013 to 2024, summing all entries within the corresponding geographic region. Figures (a), (b) and (c) are color-coded by type of disaster, as indicated in legend in (a) while Fig. (d) is color-coded following type of extreme event

The high costs associated with drought events can be attributed to the fact that Brazil is a major agricultural power and depends largely on hydroelectric plants for the majority of its energy production. Agriculture accounts for the highest total cost of the disasters, amounting to USD 42 billion. The agricultural sector contributes to around 6% of the value added to the gross domestic product (GDP) from 2011 to 2021 (IBGE, 2023). However, when considering activities such as processing and distribution, Brazil's agricultural and food sectors collectively contributed 38% of the country's GDP (CEPEA,2024) (averaged over 2013-2023). According to World Bank report (WB,2022), agriculture has the greatest impact on poverty among the four scenarios studied, as impoverished individuals are more vulnerable to food price fluctuations and depend heavily on agricultural and ecosystem-related incomes.

Figure 5d shows the distribution of three different types of EE in Brazil, as shown in the legend. Since TX90p events have increased far more than precipitation in absolute terms (up to 150% versus 30% for rainfall, Figures 2 and 3), TX90p dominates the map. However, strong rainfall events (R20mm) and droughts (DD90) are also numerous across all regions.

Notably, the figure illustrates that extreme events do not necessarily translate into disasters; a disaster results from the combination of a climate hazard and social vulnerability (Cardona et al., 2012). For example, in the Northeast, large numbers of displacements caused by heavy rains are observed, even though the frequency of R20mm is not particularly high in the same period—suggesting that vulnerability plays a significant role in disaster outcomes. In the South and Southeast, the high R20mm values align more clearly with the large numbers of people affected by storms and floods. In the Center-West, the combination of frequent heat waves (TX90p) and prolonged dry periods (DD90d) is consistent with the prevalence of disasters classified as "others," which in this context likely includes wildfires.

The analysis of extreme weather events discussed in the previous section indicates that the Northeast region has become drier over the past 30 years, a trend previously reported by Marengo et al. (2017), which could exacerbate the drought impacts shown in the panels of Fig. 5. Also, the South region has become wetter in the last 30 years, a trend that also appears in panels of Fig. 5-b,c as the most impactul type of disaster. This aligns with observed increases in heavy precipitation days (R20mm) in southern Brazil (Zilli et al., 2016), where urbanization and South Atlantic Convergence Zone dynamics may amplify extremes.

## 4 CONCLUSIONS

This study has conducted a parallel analysis of two distinct databases: the meteorological station dataset from INMET, encompassing 60 years of data recorded by over 287 conventional meteorological stations, and the S2iD database, which focuses on natural disasters, spans a 12-year period and is manually filled.

Using INMET data, the anomaly of the air temperature averaged over all stations was measured. For both the maximum and minimum daily temperature, an average increase of over 1 °C was identified in the period from 1990 to 2020 when compared to the reference period (1961 to 1990). Several climate extreme indices were then measured and showed consistently that events of maximum daily air temperature, particularly warm days (TX90p), have severely increased in comparison to the reference period, with some stations recording an increase of over 100%, while cold nights (TN10p) have decreased up to 75% in the same period. In terms of rainfall, the analysis points for a drier weather in the Northeast region and wetter weather in the South. Moreover, the number of days with heavy precipitation have increased by about 20% in the South region and decreased by approximately the same amount in the Northeast in the last three decades when compared to the reference period. Throughout the country, an increase of about 20% in extreme drought events can be seen, with the exception of the South region.

The changes in the trends of temperature and precipitation extremes are particularly important in Brazil because the country's economy is largely based on agriculture and livestock, and most of its energy is generated by hydroelectric plants. About 40% of the Brazilian GDP is related to these activities (CEPEA, 2024), making the country vulnerable to climate changes, which stresses the need for adaptation. These vulnerabilities are partly measured by the S2iD data, which shows that drought is the type of event that has the highest impact in terms of economic losses, while storms stand out as the most impactful in affecting directly people's lives, including the need of displacing them from their homes. This impact is not homogeneously felt in Brazil, reflecting the country's social inequalities.

The main limitation of this study is its inability to directly link temperature and precipitation extremes to natural disasters. Establishing connections between datasets could improve disaster prevention strategies. One approach involves applying Extreme Value Theory (EVT) (Majumdar et al., 2020) and integrating gridded meteorological data (Xavier et al.,2016) with more frequent disaster records. By using the S2iD platform to identify high-risk regions and analyzing nearby meteorological data, EVT can estimate the probability of extreme events leading to disasters. Understanding the thresholds for disaster-triggering events is essential for developing effective, region-specific adaptation measures.

## ACKNOWLEDGEMENTS

The authors are thankful to CNPq, CAPES and FAPERGS for partly funding this study. The authors have no relevant financial or non-financial interests to disclose.

## REFERENCES

Almeida, C. T., Oliveira‐Júnior, J. F., Delgado, R. C., Cubo, P., & Ramos, M. C. (2017). Spatiotemporal rainfall and temperature trends throughout the Brazilian Legal Amazon, 1973–2013. *International Journal of Climatology*, *37*(4), 2013-2026. https://doi.org/10.1002/joc.4831

Ávila, A., Justino, F., Wilson, A., Bromwich, D., & Amorim, M. (2016). Recent precipitation trends, flash floods and landslides in southern Brazil. *Environmental Research Letters*, *11*(11), 114029. https://doi.org/10.1088/1748-9326/11/11/114029

Avila-Diaz, A., Benezoli, V., Justino, F., Torres, R., & Wilson, A. (2020). Assessing current and future trends of climate extremes across Brazil based on reanalyses and earth system model projections. *Climate Dynamics*, *55*(5), 1403-1426. https://doi.org/10.1007/s00382-020-05333-z

Bador, M., Donat, M. G., Geoffroy, O., & Alexander, L. V. (2018). Assessing the robustness of future extreme precipitation intensification in the CMIP5 ensemble. *Journal of Climate*, *31*(16), 6505-6525. https://doi.org/10.1175/JCLI-D-17-0683.1

Caleffi, F., Viegas, C. V. ., Lima, K. B., & Bonato, S. V. (2024). Os impactos de eventos climáticos extremos: uma análise abrangente das enchentes de 2024 no Rio Grande do Sul. *Redes*, *29*(1), 1-27. https://doi.org/10.17058/redes.v29i1.19660

Cardona, O. D., van Aalst, M. K., Birkmann, J., Fordham, M., Mc Gregor, G., Perez, R., Pulwarty, R. S., Schipper, E. L. F. & Sihn, B. T. (2012). Determinants of risk: exposure and vulnerability. In C. B. Field et al. (Eds.), *Managing the risks of extreme events and disasters to advance climate change adaptation: special report of the intergovernmental panel on climate change* (pp. 65-108). Cambridge University Press. https://doi.org/10.1017/CBO9781139177245.005

Carvalho, I. C. D. H. (2019). *Análise de recorrências de eventos de desastres naturais com base no Sistema Integrado de Informações sobre Desastres (S2iD) e séries históricas de precipitação no Brasil: uma contribuição metodológica.* [Doctoral Dissertation, Universidade de Brasília].

Centro de Estudos Avançados em Economia Aplicada. (2024). *Cepea-USP/CNA: PIB do agronegócio brasileiro – junho de 2024.* https://www.cepea.esalq.usp.br/br/pib-do-agronegocio-brasileiro.aspx

Chagas, V. B., Chaffe, P. L., & Blöschl, G. (2022). Climate and land management accelerate the Brazilian water cycle. *Nature Communications*, *13*(1), 5136. https://doi.org/10.1038/s41467-022-32580-x

Costa, R. L., de Mello Baptista, G. M., Gomes, H. B., dos Santos Silva, F. D., da Rocha Junior, R. L., de Araújo Salvador, M., & Herdies, D. L. (2020). Analysis of climate extremes indices over northeast Brazil from 1961 to 2014. *Weather and Climate Extremes*, *28*, 100254. https://doi.org/10.1016/j.wace.2020.100254

Centre for Research on the Epidemiology of Disasters. (2022). *2021 disasters in numbers. https://www.emdat.be*

Dalagnol, R., Gramcianinov, C. B., Crespo, N. M., Luiz, R., Chiquetto, J. B., Marques, M. T., Neto, G. D., Abreu, R. C., Li, S., Lott, F. C., Anderson, L. O., & Sparrow, S. (2022). Extreme rainfall and its impacts in the Brazilian Minas Gerais state in January 2020: Can we blame climate change?. *Climate Resilience and Sustainability*, *1*(1), e15. https://doi.org/10.1002/cli2.15

Dufek, A. S., & Ambrizzi, T. (2008). Precipitation variability in São Paulo State, Brazil. *Theoretical and Applied Climatology*, *93*(3), 167-178. https://doi.org/10.1007/s00704-007-0348-7

Ebi, K. L., Vanos, J., Baldwin, J. W., Bell, J. E., Hondula, D. M., Errett, N. A., Hayes, K., Reid, C. E., Saha, S., Spector, J., & Berry, P. (2021). Extreme weather and climate change: population health and health system implications. *Annual Review of Public Health*, *42*(1), 293-315. https://doi.org/10.1146/annurev-publhealth-012420-105026

Governo do Estado do Rio Grande do Sul (2024). *Boletins sobre o impacto das chuvas no RS.* https://www.estado.rs.gov.br/boletins-sobre-o-impacto-das-chuvas-no-rs

Hänsel, S., Schucknecht, A., & Matschullat, J. (2016). The Modified Rainfall Anomaly Index (mRAI)—is this an alternative to the Standardised Precipitation Index (SPI) in evaluating future extreme precipitation characteristics?. *Theoretical and Applied Climatology*, *123*(3), 827-844. https://doi.org/10.1007/s00704-015-1389-y

Haylock, M. R., Peterson, T. C., Alves, L. M., Ambrizzi, T., Anunciação, Y. M. T., Baez, J., Barros, V. R., Berlato, M. A., Bidegain, M., Coronel, G., Corradi, V., Garcia, V.J., Grimm, A. M., Karoly, D., Marengo, J. A., Marino, M. B., Moncunill, D. F., Nechet, D., Quintana, J., Rebello, E., Rusticucci, M., Santos, J. L., Trebejo, I., Vincent, L. A. (2006). Trends in total and extreme South American rainfall in 1960–2000 and links with sea surface temperature. *Journal of Climate*, *19*(8), 1490-1512. https://doi.org/10.1175/JCLI3695.1

Instituto Brasileiro de Geografia e Estatística (IBGE). (2023). *Contas nacionais trimestrais: indicadores de volume e valores correntes*. https://agenciadenoticias.ibge.gov.br/agencia-sala-de-imprensa/2013-agencia-de-noticias/releases/39303-pib-cresce-2-9-em-2023-e-fecha-o-ano-em-r-10-9-trilhoes

Instituto Brasileiro de Geografia e Estatística (IBGE). (2022). *Censo demográfico 2022*. IBGE.

Medeiros, F. J., de Oliveira, C. P., & Avila-Diaz, A. (2022). Evaluation of extreme precipitation climate indices and their projected changes for Brazil: From CMIP3 to CMIP6. *Weather and Climate Extremes*, *38*, 100511. https://doi.org/10.1016/j.wace.2022.100511

Kuhn, C. E., Reis, F. A., Oliveira, V. G., Cabral, V. C., Gabelini, B. M., & Veloso, V. Q. (2022). Evolution of public policies on natural disasters in brazil and worldwide. *Anais da Academia Brasileira de Ciências*, *94*, e20210869. https://doi.org/10.1590/0001-3765202220210869

de Lima, J. A. G., & Alcântara, C. R. (2019). Comparison between ERA Interim/ECMWF, CFSR, NCEP/NCAR reanalysis, and observational datasets over the eastern part of the Brazilian Northeast Region. *Theoretical and Applied Climatology*, *138*(3), 2021-2041. https://doi.org/10.1007/s00704-019-02921-w

Lagos-Zúñiga, M., Balmaceda-Huarte, R., Regoto, P., Torrez, P., Olmo, M., Lyra, A., Quispe-Pareja, D., & Bertolli, M. L. (2024). Extreme indices of temperature and precipitation in South America: Trends and intercomparison of regional climate models. *Climate Dynamics*, 62, 4541–4562. https://doi.org/10.1007/s00382-022-06598-2

Luiz-Silva, W., & Oscar-Júnior, A. C. (2022). Climate extremes related with rainfall in the State of Rio de Janeiro, Brazil: A review of climatological characteristics and recorded trends. *Natural Hazards*, 114(1), 713–732. https://doi.org/10.1007/s11069-022-05409-5

Majumdar, S. N., Pal, A., & Schehr, G. (2020). Extreme value statistics of correlated random variables: a pedagogical review. *Physics Reports*, *840*, 1-32. https://doi.org/10.1016/j.physrep.2019.10.005

Mann, H. B. (1945). Nonparametric tests against trend. *Econometrica: Journal of the Econometric Society*, 245-259. https://doi.org/10.2307/1907187

Marengo, J. A., Torres, R. R., & Alves, L. M. (2017). Drought in Northeast Brazil—past, present, and future. *Theoretical and Applied Climatology*, *129*(3), 1189-1200. https://doi.org/10.1007/s00704-016-1840-8

Masson-Delmotte, V., Zhai, P., Pirani, A., Connors, S. L., Pean, C., Berger, S., Caud, N., Chen, Y., Goldfarb, L., Gomis, M. I., Huang, M., Leitzell, K., Lonnoy, E., Matthews, J. B. R., Maycock, T. K., Waterfield, T., Yelekci, O., Yu, R., & Zhou, B. (2021). Climate change 2021: the physical science basis. *Contribution of working group I to the sixth assessment report of the intergovernmental panel on climate change*, *2*(1), 2391. https://www.ipcc.ch/report/ar6/wg1/downloads/report/IPCC_AR6_WGI_FrontMatter.pdf

McKee, T. B., Doesken, N. J., & Kleist, J. (1993. The relationship of drought frequency and duration to time scales. *In Proceedings of the 8th Conference on Applied Climatology* (pp. 179-183). *American Meteorological Society.* https://climate.colostate.edu/pdfs/relationshipofdroughtfrequency.pdf

de los Milagros Skansi, M., Brunet, M., Sigró, J., Aguilar, E., Arevalo Groening, J. A., Bentancur, O. J., Castellón Geier, Y. R., Correa Amaya, R. L., Jácome, H., Malheiros Ramos, A., Oria Rojas, C., Pasten, A. M., Sallons Mitro, S., Villaroel Jiménez, C., Martínez, R., Alexander, L. V., Jones, P. (2013). Warming and wetting signals emerging from analysis of changes in climate extreme indices over South America. *Global and Planetary Change*, *100*, 295-307. https://doi.org/10.1016/j.gloplacha.2012.11.004

Minervino, A. C., & Duarte, E. C. (2016). Danos materiais causados à Saúde Pública e à sociedade decorrentes de inundações e enxurradas no Brasil, 2010-2014: dados originados dos sistemas de informação global e nacional. *Ciência & Saúde Coletiva*, *21*(3), 685-694. https://doi.org/10.1590/1413-81232015213.19922015

Murara, P., Acquaotta, F., Garzena, D., & Fratianni, S. (2019). Daily precipitation extremes and their variations in the Itajaí River Basin, Brazil. *Meteorology and Atmospheric Physics*, *131*(4), 1145-1156. https://doi.org/10.1007/s00703-018-0627-0

Otto, F. E. (2017). Attribution of weather and climate events. *Annual Review of Environment and Resources*, *42*, 627-646. https://doi.org/10.1146/annurev-environ-102016-060847

Ramos Filho, G. M. (2021). *Performance of rainfall threshold for flood identification from ground- and satellite-based (sub) daily data.* [Doctoral Dissertation, Universidade Federal da Paraíba]. https://repositorio.ufpb.br/jspui/handle/123456789/22729

Reboita, M. S., Mattos, E. V., Capucin, B. C., Souza, D. O. d., & Ferreira, G. W. d. S. (2024). A multi-scale analysis of the extreme precipitation in southern Brazil in April/May 2024. *Atmosphere*, *15*(9), 1123.https://doi.org/10.3390/atmos15091123

Regoto, P., Dereczynski, C., Chou, S. C., & Bazzanela, A. C. (2021). Observed changes in air temperature and precipitation extremes over Brazil. *International Journal of Climatology*, *41*(11), 5125-5142. https://doi.org/10.1002/joc.7119

Rohde, R. A., & Hausfather, Z. (2020). The Berkeley Earth land/ocean temperature record. *Earth System Science Data*, *2020*, 1-16. https://doi.org/10.5194/essd-12-3469-2020

Rooy, M. (1965). A rainfall anomaly index independent of time and space. *Notos*, 14, 43–48.

Rosso, F. V., Boiaski, N. T., Ferraz, S. E., Dewes, C. F., & Tatsch, J. D. (2015). Trends and decadal variability in air temperature over Southern Brazil. *American Journal of Environmental Engineering*, *5*(1A), 85-95.

Schossler, V., Simões, J. C., Aquino, F. E., & Viana, D. R. (2018). Precipitation anomalies in the Brazilian southern coast related to the SAM and ENSO climate variability modes. *RBRH*, *23*, e14. https://doi.org/10.1590/2318-0331.231820170081

Sen, P. K. (1968). Estimates of the regression coefficient based on Kendall's tau. *Journal of the American Statistical Association*, *63*(324), 1379-1389. https://doi.org/10.1080/01621459.1968.10480934

Silva Dias, M. A., Dias, J., Carvalho, L. M., Freitas, E. D., & Silva Dias, P. L. (2013). Changes in extreme daily rainfall for São Paulo, Brazil. *Climatic Change*, *116*(3), 705-722. https://doi.org/10.1007/s10584-012-0504-7

Sippel, S., Zscheischler, J., Heimann, M., Otto, F. E., Peters, J., & Mahecha, M. D. (2015). Quantifying changes in climate variability and extremes: Pitfalls and their overcoming. *Geophysical Research Letters*, *42*(22), 9990-9998. https://doi.org/10.1002/2015GL066307

Solberg, K. (2010). Worst floods in living memory leave Pakistan in paralysis. *The Lancet (London, England)*, *376*(9746), 1039-1040. https://doi.org/10.1016/s0140-6736(10)61469-9

Svoboda, M., Hayes, M., Wood, D. (2012). Standardized precipitation index: user guide. (WMO-No. 1090). *World Meteorological Organization.*

Tank, A., Zwiers, F., Zhang, X. (2009). Guidelines on analysis of extremes in a changing climate in support of informed decisions for adaptation. World Meteorological Organization.

Instituto Nacional de Meteorologia. (2022). Banco de dados meteorológicos: Normais climatológicas (1961–1990). https://portal.inmet.gov.br/

Governo Federal do Brasil. (2025). Sistema integrado de informações sobre desastres (S2iD). Ministério da Integração e do Desenvolvimento Regional. http://S2iD.mi.gov.br/

Vicente-Serrano, S. M., Beguería, S., & López-Moreno, J. I. (2010). A multiscalar drought index sensitive to global warming: the standardized precipitation evapotranspiration index. *Journal of climate*, *23*(7), 1696-1718. https://doi.org/10.1175/2009JCLI2909.1

World Bank. (2022). The World Bank annual report 2022: Helping countries adapt to a changing world. World Bank Group. http://documents.worldbank.org/curated/en/099030009272214630

World Meteorological Organization. (2024). State of the climate 2024 (WMO-No. 1300). https://library.wmo.int/records/item/69075-state-of-the-climate-2024

Xavier, A. C., King, C. W., & Scanlon, B. R. (2016). Daily gridded meteorological variables in Brazil (1980–2013). *International Journal of Climatology*, 36(6), 2644-2659. https://doi.org/10.1002/joc.4518

Zhang, X., & Yang, F. (2004). RClimDex (1.0) user manual. *Climate Research Branch Environment Canada*, *22*, 13-14. https://rcc.acmad.org/procedure/RClimDexUserManual.pdf

Zhang, D. L., Lin, Y., Zhao, P., Yu, X., Wang, S., Kang, H., & Ding, Y. (2013). The Beijing extreme rainfall of 21 July 2012: “Right results” but for wrong reasons. *Geophysical Research Letters*, *40*(7), 1426-1431. https://doi.org/10.1002/grl.50304

Zhang, X., Hegerl, G., Zwiers, F. W., & Kenyon, J. (2005). Avoiding inhomogeneity in percentile-based indices of temperature extremes. *Journal of Climate*, *18*(11), 1641-1651. https://doi.org/10.1175/JCLI3366.1

Zhang, X., Alexander, L., Hegerl, G. C., Jones, P., Tank, A. K., Peterson, T. C., Trewin, B., & Zwiers, F. W. (2011). Indices for monitoring changes in extremes based on daily temperature and precipitation data. Wiley Interdisciplinary Reviews: *Climate Change*, 2(6), 851-870. https://doi.org/10.1002/wcc.147

Zilli, M. T., Carvalho, L. M., Liebmann, B., & Silva Dias, M. A. (2017). A comprehensive analysis of trends in extreme precipitation over southeastern coast of Brazil. *International Journal of Climatology*, 37(5), 2269-2279. https://doi.org/10.1002/joc.4840

# ANNEX – SUPPLEMENTARY MATERIAL (CLIMATIC EXTREMES IN BRAZIL: A PARALLEL ANALYSIS OF HISTORICAL TRENDS AND SOCIOECONOMICAL IMPACTS)

## METHODS

In this section we provide complementary information on the databases used, show the data robustness tests and additional temperature and precipitation indices.

### S2iD Base

The S2iD database is the official resource when a city declares a state of calamity or catastrophe in Brazil. It is the first official dataset to be filled with information, which is later confirmed by authorities. A drawback of this database is that it is not automated and is manually filled, which can introduce errors. To address this issue, we cleaned events that were spatially and temporally close to each other, as it happened that the same event was registered twice. Additionally, in 2023, one event reported almost USD 130 billion in damages and was discarded, as this value is absurdly high compared to other EE values and the Brazilian GDP itself, and lastly we just considered events with status Confirmed, what means that they were confirmed by Brazilian authorities.

As mentioned in the main text, the S2iD database contains several entries to identify different types of disasters. In this work, we group all these types into five categories as shown in Table 1.

Figure 1 – This table summarizes how we group different types of disasters. The left side of the table shows our classification, while the right side specifies the types of events that are registered in the S2iD

| Classification | Original name in S2iD base |
|---|---|
| **Storm** | Heavy Rain |
| | Hail |
| | Tornado |
| | Gale |
| | Lightning Storm |
| **Drought** | Dry |
| | Drought |
| **Flooding** | Flood |
| | Overflows |
| | Flash Flood |
| **Disease** | Bacterial, Fungal, Parasitic and Viral Infectious Diseases |
| **Other** | Wildfires/Forest Fires |
| | Heat Wave or Cold |
| | Mass movement |
| | Landslide |
| | Cyclones |
| | Another 40 types of less frequent events |

An interesting aspect of this database is that certain locations appear multiple times, as illustrated in Fig. 2. This may be because local authorities learn how to use the database and proceed to include every disaster that happens in their city.

Figure 2 – Sum of extreme occurrences per City and State in the time from 2013-2021 from the S2iD basis

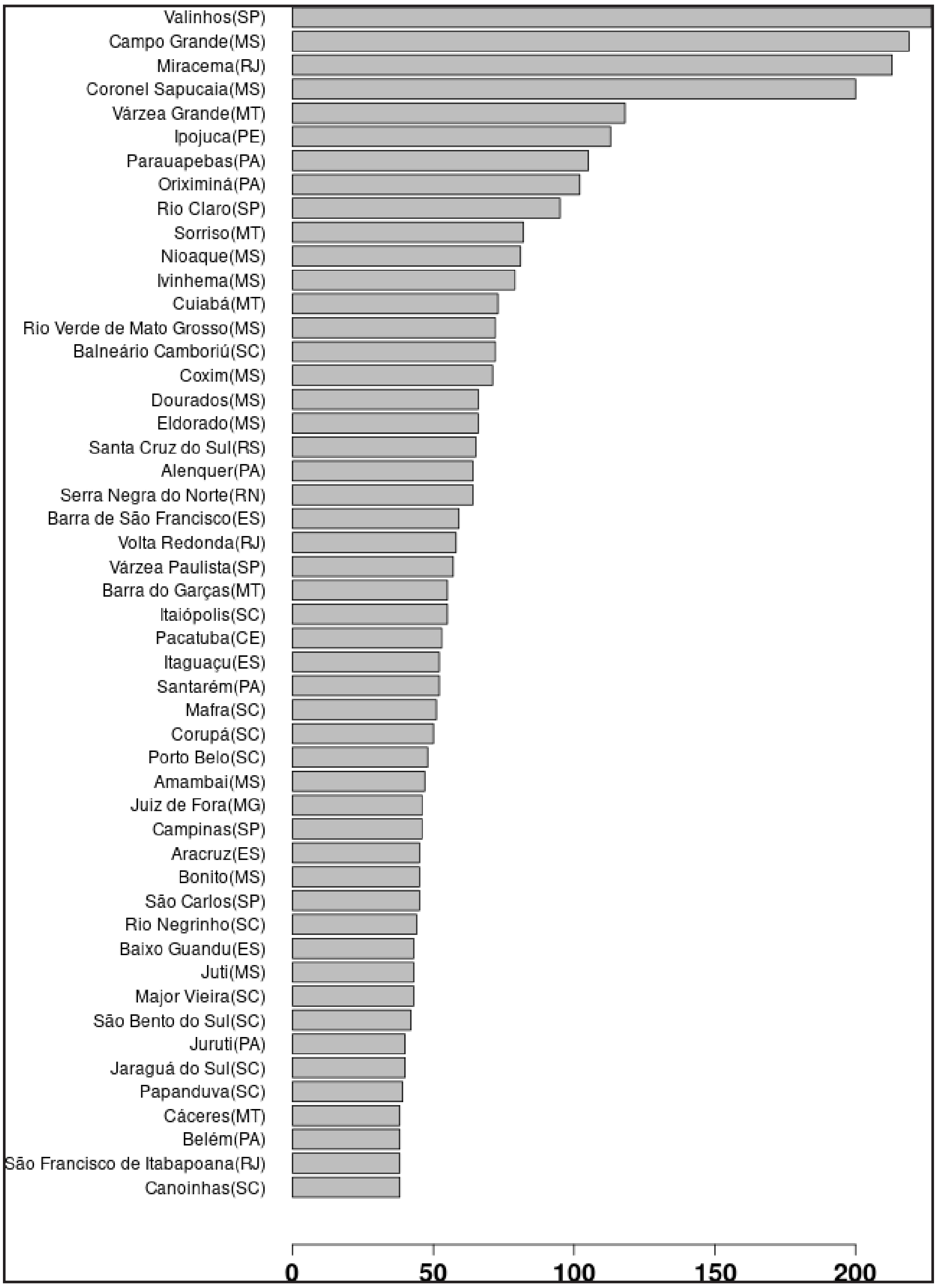

## Exclusion of INMET meteorological stations and outliers

As mentioned in the main text, we excluded stations with less than 30% of data in the RP (1961-1990) and the whole period of analysis (1961-2020), as well as individual years where a station worked for less than 30% of the time, which resulted in approximately 287 working stations. In the next section we present some of the temperature indices calculated and how the different exclusion criteria based on percentage of working time change Extreme Event (EE) occurrences (Figure 5). Our choice of 30% as the working time exclusion threshold is then justified as an average that retains the general trend of different threshold curves while keeping a reasonably good amount of data.

To define data outliers we consider the standard definitions used by Climpact quality control, where we discard temperature data lower than p25-3*IQR (percentile 25 minus 3 times the interquartile range of the temperature distribution), upper than p75+3*IQR and daily precipitation volume data above p75+5*IQR.

## 2 IDENTIFICATION OF EXTREME EVENTS

### Temperature Indices

The TN10p and TX90p are indices that reference the tails of the minimum and maximum daily air temperature distributions, as described in [7, 4]. They relate to the temperatures under the 10% and over the 90% percentiles of daily minimum (TN) and maximum (TX) air temperatures, respectively. In this work, these percentiles are calculated with data only in the reference period to define a percentile based extreme line for data in the whole period. This way, the identification of EE in the analysis period takes into account what was extreme in the reference period. The same is considered for the TN90p and TX10p, which relate to the temperatures over the 90% and under the 10% percentiles of TN and TX temperatures, respectively.

Figure 3 – TN10p and TX90p EE detection threshold in a station in Montes Claros, in the state of Minas Gerais. The x axis represents days of the year (1-366), and the dots in both figures represent daily temperatures in all of the WP. The cyan and red lines in the left and right figures respectively represent the 10th and 90th data percentile, which are used as the extreme threshold. Blue points in the left figure identify minimum temperature extremes (TN10p). Red points in the right figure represent maximum temperature extremes

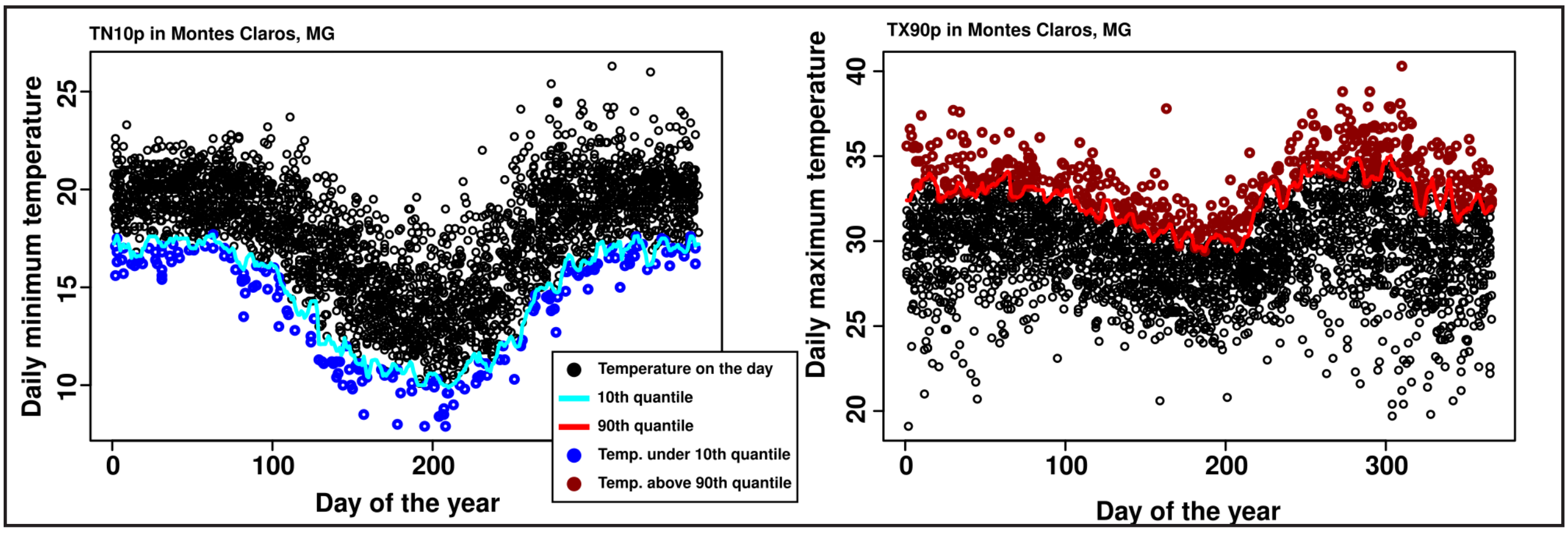

In Fig. 3 TN (left) and TX (right) time series are shown in the year-day representation for the RP, with the daily temperature data organized by day of the year (1 to 365/366). The figure helps to show how the indices account for seasonal effects, and how for each day of the year there is a different percentile threshold. This moving threshold is calculated using a 5-day window to increase statistics: let us suppose that day *j* has 30 pieces of data on the TX distribution for the RP, considering the window [j-2,j+2] we increase this number to 150 pieces, which is then used to define a temperature threshold using the 90% percentile - for instance, 36ºC for the 90% percentile, so every TX > 36ºC for day *j* will be considered as extreme. The extreme temperature events are represented as blue points, under the blue line, for TN10p (on the left), and as red points, above the red line, for TX90p (on the right).

Figure 4 – Both figures represent temperature series varying the extreme percentile threshold. On the left, TX is shown and the three lines represent the use of 90%, 95% and 99% as thresholds. The same is done on the left using the TN and the 10%, 5% and 1% thresholds

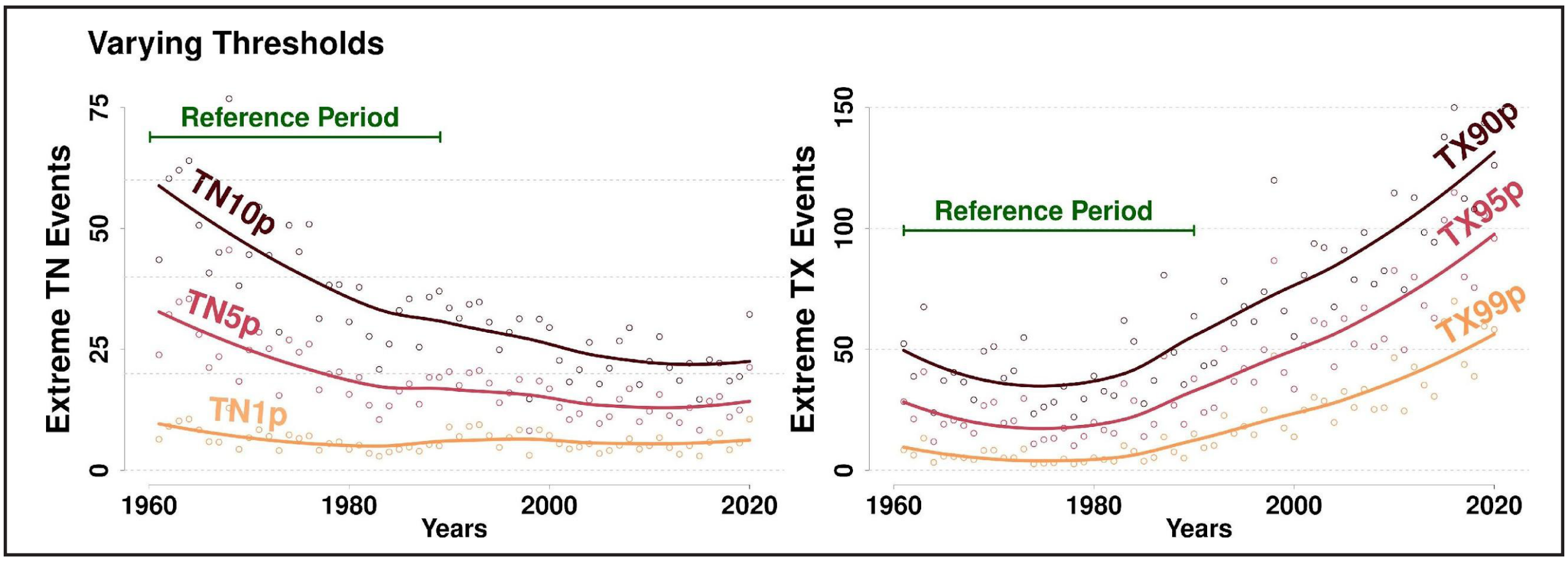

Figure 4 shows distinct time series for more rigorous extreme event thresholds (TN < 1%, 5% and 10% percentiles and TX > 99%, 95% and 90% percentiles). We see that the extreme TN events decay while extreme TX events increase independently of the percentile threshold chosen. Similarly, these results are robust for even less rigorous thresholds. As expected, the more rigorous the extreme event threshold chosen, the less events identified. Considering this, we use the 10% and 90% percentile thresholds for the minimum and maximum temperature analysis, respectively.

As mentioned in Section 1, we excluded stations which worked for less than 30% of the whole period and the RP from the whole analysis. Additionally, individual years where a station worked for less than the same 30% were not considered. To choose this fraction threshold, we tested cuts from 10-50%, considering that different cuts could imply changes in the EE identification. In figure 5, extreme TN and TX time series are presented showing how each cut influences the data behavior. In the *Other Precipitation Indices* section, the same tests are shown for precipitation indices. Since the general shape of the curves do not change much, in order to maintain a balance between the amount of data and the robustness of our results, the 30% threshold was chosen.

Figure 5 – TX90p and TN10p EE detection varying station functioning cut. Each line and points are separated by percentage cut and identified in a particular color. The figures on the left and right represent the TN and TX behavior

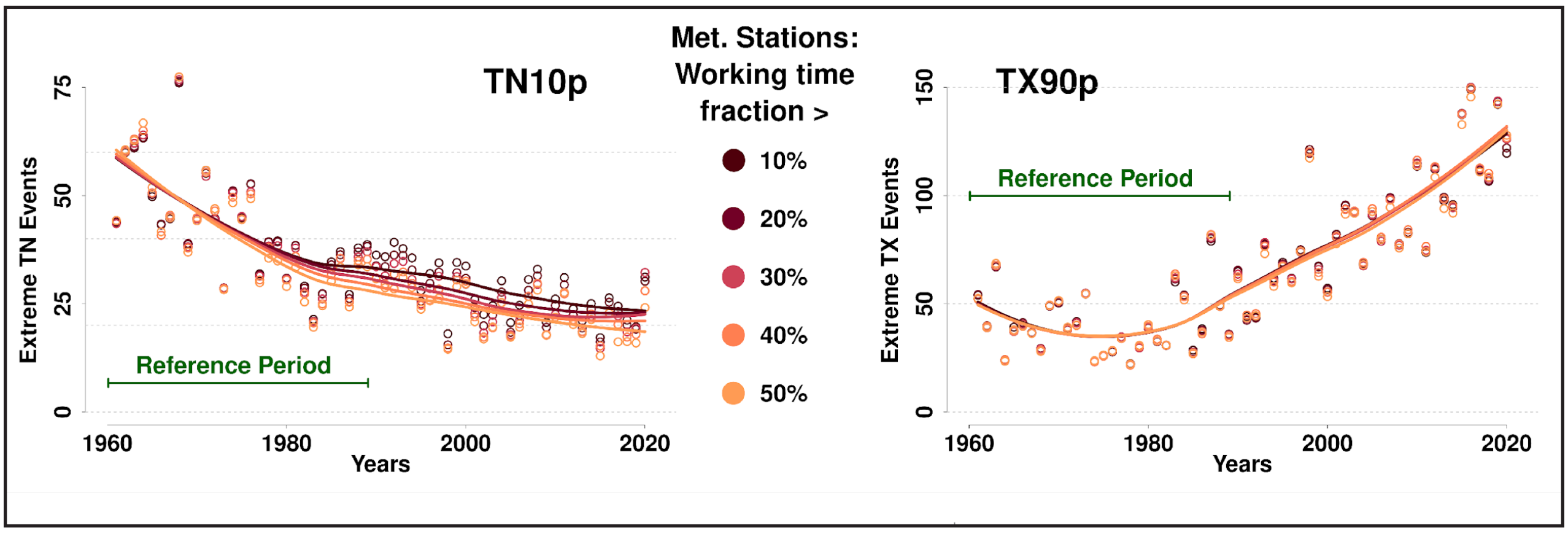

## Precipitation Indices: SPI and SPEI

The precipitation indices used are the Standard Precipitation Index (SPI) [2] and the Standard Precipitation Evapotranspiration Index (SPEI) [6]. The SPI considers only precipitation data, while the SPEI also takes temperature into account. The indices SPI3, SPEI3, SPI6, SPEI6, SPI12, and SPEI12 are calculated using the Climpact software (https://climpact-sci.org/). These indices were not used to the evaluation of extremes (even that they could be), but just to the trend definition of the time series obtained, in order to understand the overall behavior of precipitation in each region. The trend for those series of self-correlation are tested by the Mann-Kendall test for in-homogeneous data, that is a median over possible coefficients that do not overestimate the extremes [1], if the series have enough and sufficiently continuous data, they pass the MK test, and the Sen's slope trend [3] is calculated. These trends show the tendency that each station have to become drier or wetter, depending on the time scale of each index.

The SPI indices calculated through Climpact software are obtained exactly as stated by McKee et al. [2]: monthly precipitation data are collected for a period of at least 360 months. A set of averaging periods *j* months is chosen, where *j* = 3, 6, 12 months, representing typical time scales for precipitation deficits affecting water sources. Moving datasets are generated such that each month a new value is calculated using precipitation data from the previous *i* months, where *i* = *j*. Each dataset is fitted to a Gamma distribution to establish the probability-to-precipitation relationship. The historic Gamma distribution relationship is then used to compute the probability of any observed precipitation data point, followed by the application of the inverse normal transformation to the computed probabilities. The precipitation deviation is subsequently converted into a normally distributed value with a mean of zero and a standard deviation of one. The resulting value from the normal distribution transformation represents the SPI for the specific precipitation data point.

To calculate the SPEI indices, the Climpact software perform a similar algorithm that for SPI, but with some differences: it calculates the potential evapotranspiration (PET) through Thornthwaite method [5], then calculate the climatic balance for each month, as the difference between precipitation and PET. For each month is calculated the aggregated values over the previous *n* months (where *n* = 3, 6, 12). The biggest difference from SPI index is that for the SPEI a log-logistic distribution is used to fit the aggregated values, which need the fitting of 3 moments for the distribution. After the distribution is obtained, the software transforms the cumulative probability to a standard normal distribution and this represents the SPEI for the specific data point.

Figure 6 presents an example of the plots generated by the Climpact software for Station 20 (Óbidos, PA in the North region). The y-axis indicates the magnitude of autocorrelation and the x-axis represents the years in the series. The upper graphs display SPI3 (left) and SPEI3 (right), both with the respective Sen’s slope trends. The lower graphs show SPI12 (left) and SPEI12 (right), which lack sufficient data to produce the Sen’s slope trends.

Figure 6 – Examples of SPE and SPEI indexes for the metereological station 20 (Obidos, PA on the North region). The upper graphs display SPI3 (left) and SPEI3 (right), both with the respective Sen's slope trends. The lower graphs show SPI12 (left) and SPEI12 (right), which lack sufficient data to produce the Sen's slope trends. The y-axis represents the magnitude of autocorrelation for the precipitation series within the respective time windows (3 months for the upper panels and 12 months for the lower panels). The x-axis represents the years in the time series. Blue values indicate wet periods at this station, while red values indicate dry periods

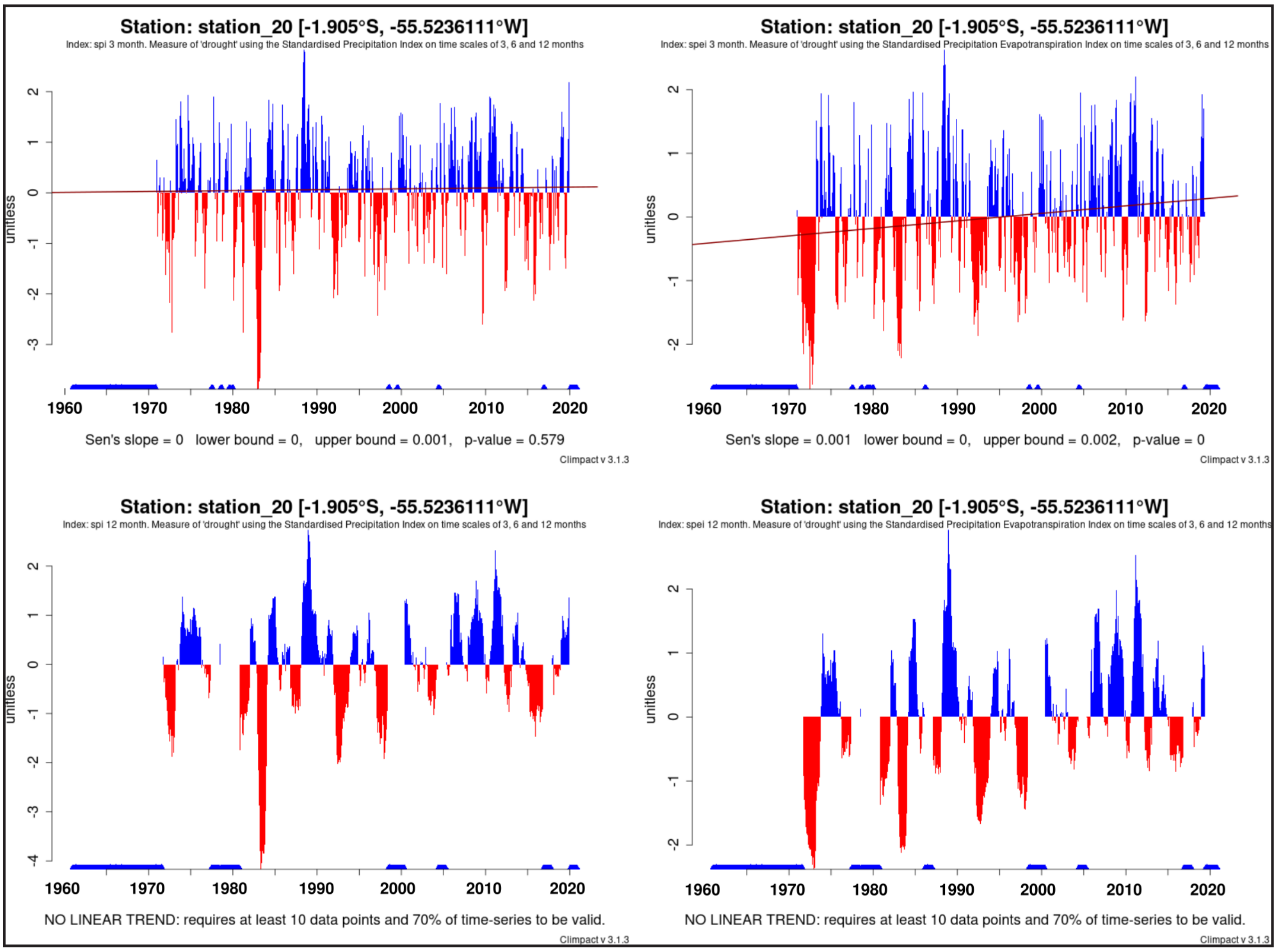

## Other Precipitation Indices

The DD90d and Rain events are indices that reference precipitation distributions. The DD90d extreme events are calculated using series of days without rain. In each station, the days with 0mm of precipitation are selected and days in a row are grouped.

We then take the 90% threshold of this distribution (using only data in the RP) and select the day series over this percentile (in the whole period), characterized as extreme DD90d events. The Rain events (*RXXmm*) are the days where a station registered over 20, 50 or 100mm of rain (*R*20*mm*,*R*50*mm* and *R*100*mm* respectively). These indices and the identification of extremes are both exemplified in figure 7, in an individual station in Montes Claros, Minas Gerais.

Figure 7 – Dry days and Rain EE detection threshold in a station in Montes Claros, in the state of Minas Gerais. The figure on the left represent the Dry Days index, where the red line represents the number of days that mark the 90% threshold in that station, and the red blocks represent the extreme events. On the right, the Rain index is represented, with the blue line representing 50mm of rain, used as this index's threshold, and blue blocks represent the extremes events

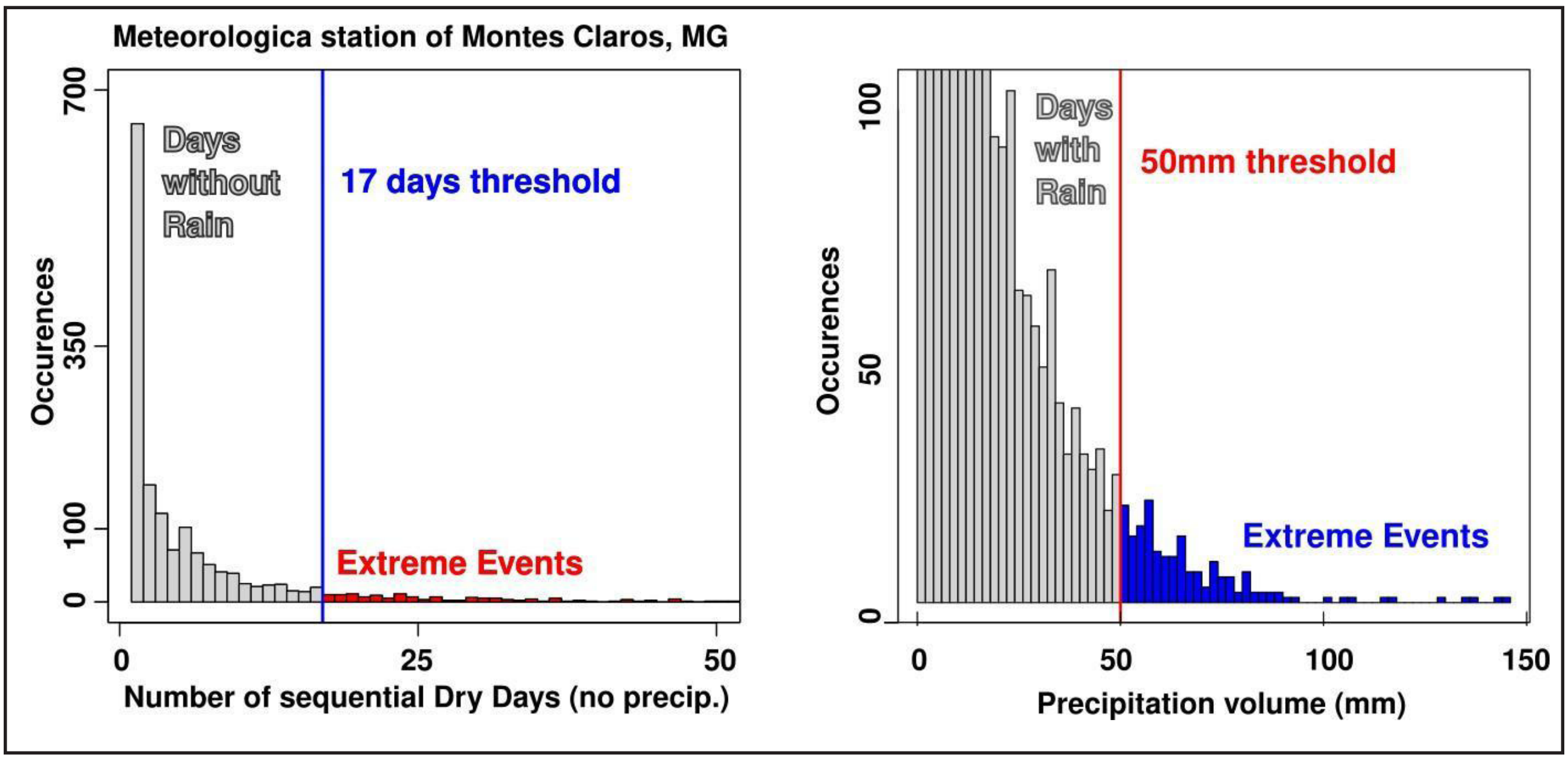

Like what was done in the temperature analysis, we tested station functioning cuts from 10-50% for precipitation indices. Similarly, all threshold cuts showed a robust behavior - shown in Fig. 8 for the R20mm (left) and DD90 (right) indices, respectively. Additionally, Figure 9 shows the 10% and 50% threshold cuts. The similarity between both maps, with the biggest difference being the amount of stations, attests to the data robustness.

Figure 8 – R20mm (left) and DD90 (right) EE detection varying station functioning cut. Each line and points are separated by percentage cut and identified in a particular color. The shaded section refers to the RP

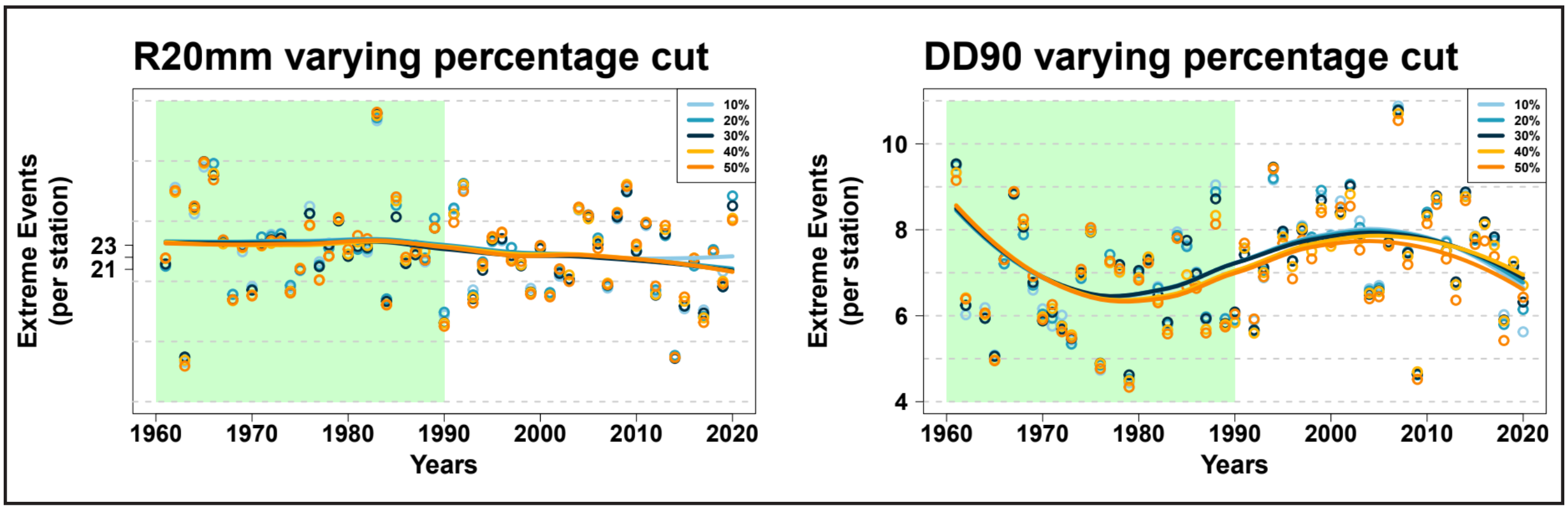

Figure 9 – Comparison of Different Thresholds for Meteorological Station Cuts. The R20mm index is analyzed across all meteorological stations in Brazil. The left panel uses a 10% threshold cut, and the right panel also uses a 50% threshold cut. The results are nearly identical, but the lower threshold excludes fewer stations, resulting in a denser and more populated figure

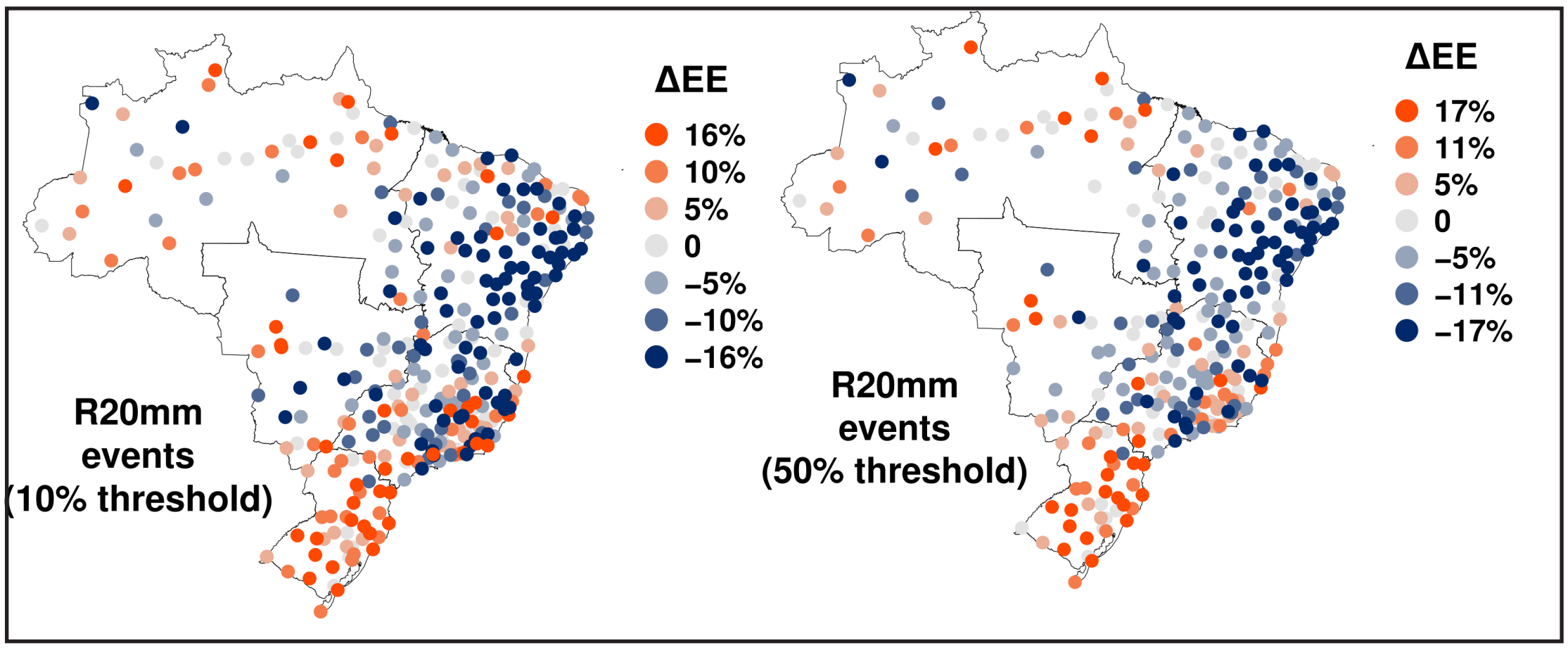

### Measure of the Mean of an Extreme Event

To calculate the mean number of EE in a given period, such as the RP or AP, we first measure the number of EE in a given year *y* and station *s*, $n_{y,s}^{EE}$ . To take into account the fact that each meteorological station does not operate the entire year, we normalize it as follows:

(1) $$EE_{y,s} = \frac{n_{y,s}^{EE}}{f_{y,s}},$$

where $f_{y,s}$ is the fraction of functioning days of each station *s* in each year *y*.

To compute the time series of EE averaged over several meteorological stations, we measure the average number of EE in a given year *y* using the following procedure:

(2) $$\overline{EE}_y = \frac{\sum_s EE_{y,s}}{f'_y},$$

where the sum is made over a range of desired meteorological stations (for example we sum up over all the stations of a Brazilian region). To take into account the fact that some stations stop working sometimes, we normalize it by $f'_y$ , which is the umber of functioning stations in a given year y. Eq.(2) is used in all temperature and precipitation time series.

To build maps such as those presented in Figs. (2) and (3) of the main text, we need to look at individual meteorological stations. To do so, we sum *EEy,s* over an interval Δy and divide it by the same interval (therefore calculating the mean EE over said period):

(3) $$\overline{EE}_{\Delta y,s} = \frac{\sum_{y \in \Delta y} EE_{y,s}}{\Delta y}.$$

In this work Eq. (3) is used when Δy is the interval of time corresponding the AP and RP, defining the quantities used in Eq.(1) of the main text:

(4) $$\overline{EE}_{RP,s} = \frac{\sum_{y\ \in\ RP} EE_{y,s}}{RP},$$

(5) $$\overline{EE}_{AP,s} = \frac{\sum_{y\ \in\ AP} EE_{y,s}}{AP},$$

$y \in RP$ ($y \in AP$) indicates that the sum is over the years when station $s$ was functioning during the reference period, RP (analysis period, AP).

## 3 OTHER RESULTS

In this section we show additional measures to test the robustness of our results and complement the data shown in the main text.

### Data on Temperature

Figure 10 shows the monthly temperature anomaly for all the working stations in Brazil, with each line representing an average over 5 years. The minimum temperature anomaly appears on the left and the maximum on the right, color coded from red (year 1960) to purple (year 2020). It is interesting to notice how the difference in temperature anomaly is much more pronounced in winter months (June-August), with an increase of around 4ºC for the TN anomaly and 3ºC for TX anomaly in the last sixty years, than the summer months (December-February) that have subtler increases of around 1ºC for both temperature series in the last sixty years.

Figure 10 – Month anomaly 5 year average in Brazil: left) Minimum temperature anomaly and right) Maximum temperature anomaly. Color code: 5 year average from 1960 (red line) to 2020 (purple line)

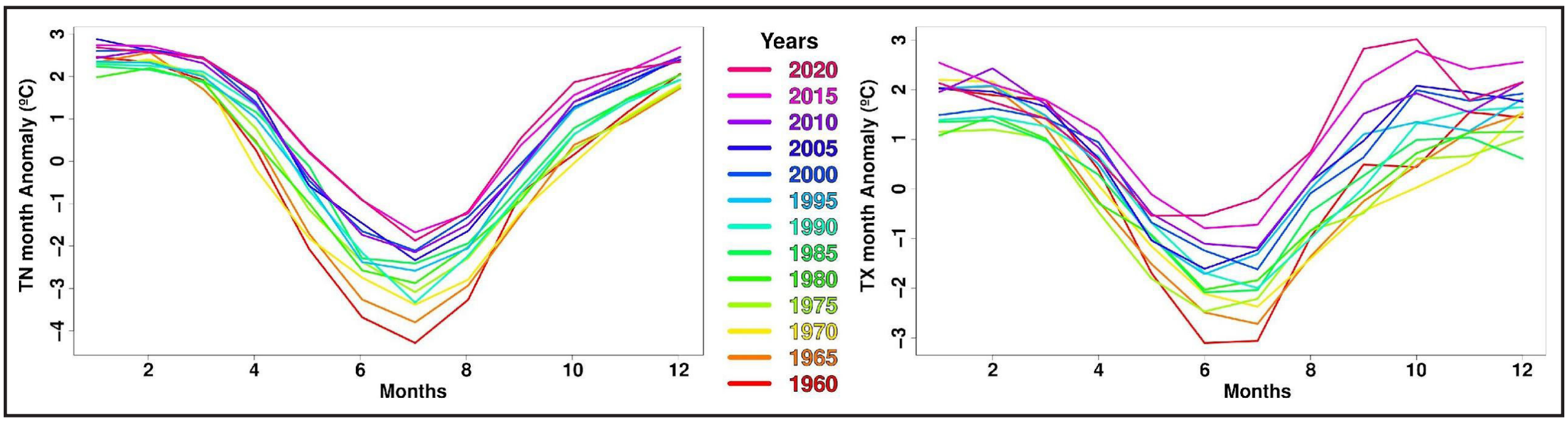

In addition to observing yearly temperature changes, we analysed the DTR, which measures the daily temperature range. The results over all meteorological stations are shown in ig. 11, where a decrease in the daily range can be noted in the RP.

Figure 11 – DTR as a time series over all stations. The red line is a smooth of the black line and shows a decrease in events in the RP, which then stabilizes in the AP

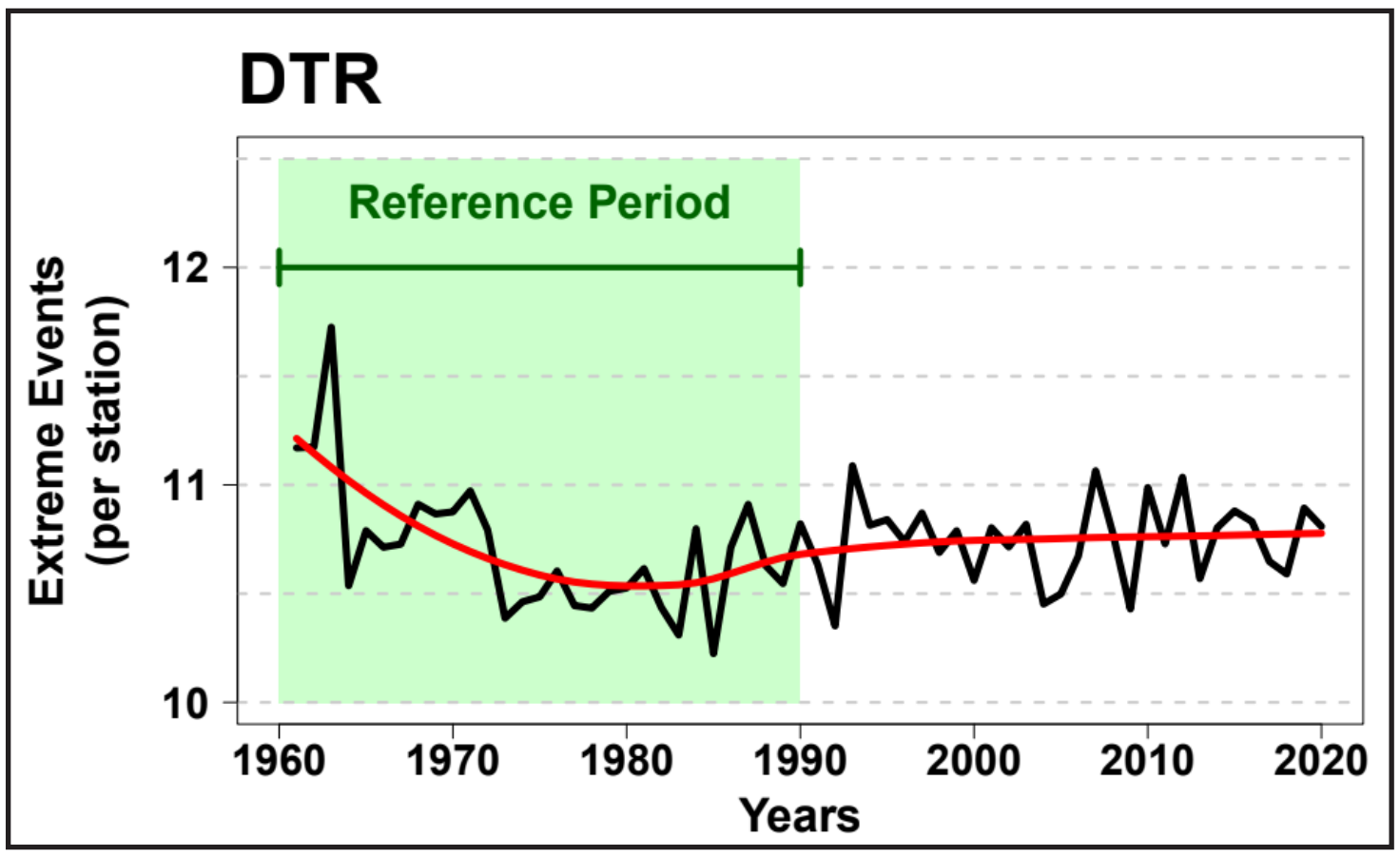

To test the robustness of our findings, we have investigated indices more closely related to cold/heat waves, that take into account the temporal continuity of an extreme event. In Figure 12 we present the average of EE that have happened *n* consecutive days: on the Fig. 12-left for the minimum temperature EE, TN10p*n*d, and on the Fig. 12-right for the maximum temperature EE, TX90p*n*d, for *n* = 1, 3 and 5. We see that the differences in EE consecutiveness do not change much the overall shape of the curves, but decrease their occurrences, as we should expect from more rare (or longer) events. We continue to see this behavior in Fig. 13, where the WSDI time series is calculated. This additional index is nothing more than an extreme maximum temperature event over 6 days (equivalent to TX90p6d), and so it comes as no surprise to see that it shows a similar behavior to the curves in Fig. 12-right.

Figure 12 – Extreme temperature events by station, for minimum temperature on the left, and maximum temperature on the right. Color-coded for the continuity of events occurrence in days, from wine (1 day) to orange (5 days)

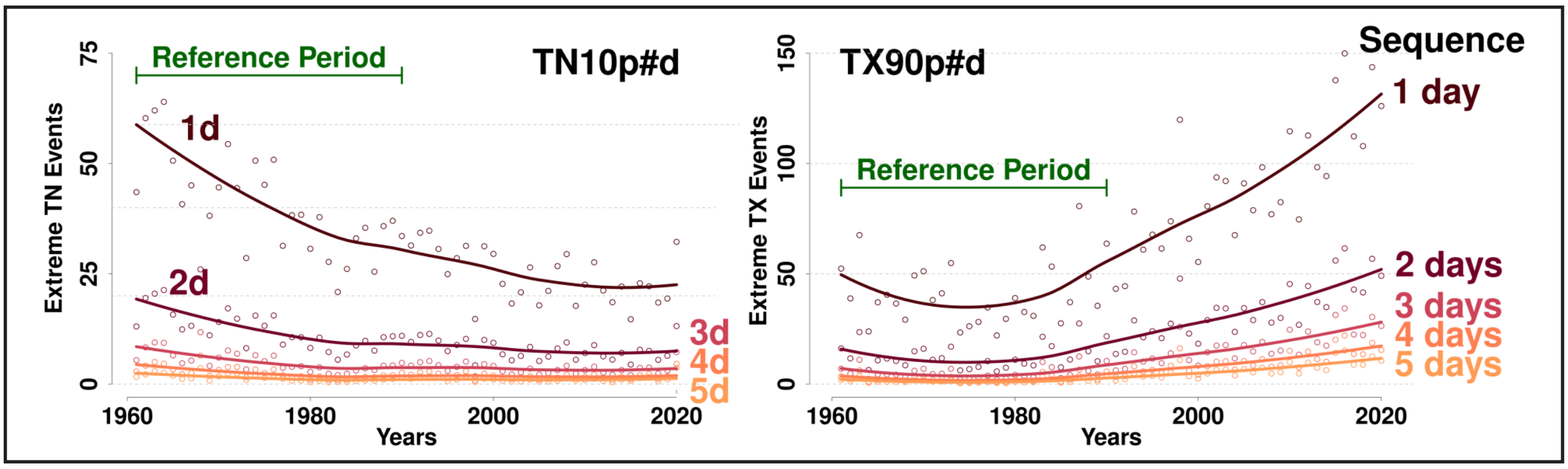

Finally, in addition to calculating the TX90p and TN10p indices, we decided to measure the TN90p and TX10p indices, seen in Figures 14 and 15, respectively. These figures are complementary to Fig. 2 in the main text, and show a similar behavior of increasing maximum temperature events (TN90p and TX90p) and decreasing minimum temperature events (TX10p and TN10p).

Figure 13 – WSDI events as a time series over all stations. The red line is a smooth of the black line and shows a clear increase in events in the AP in comparison to the RP

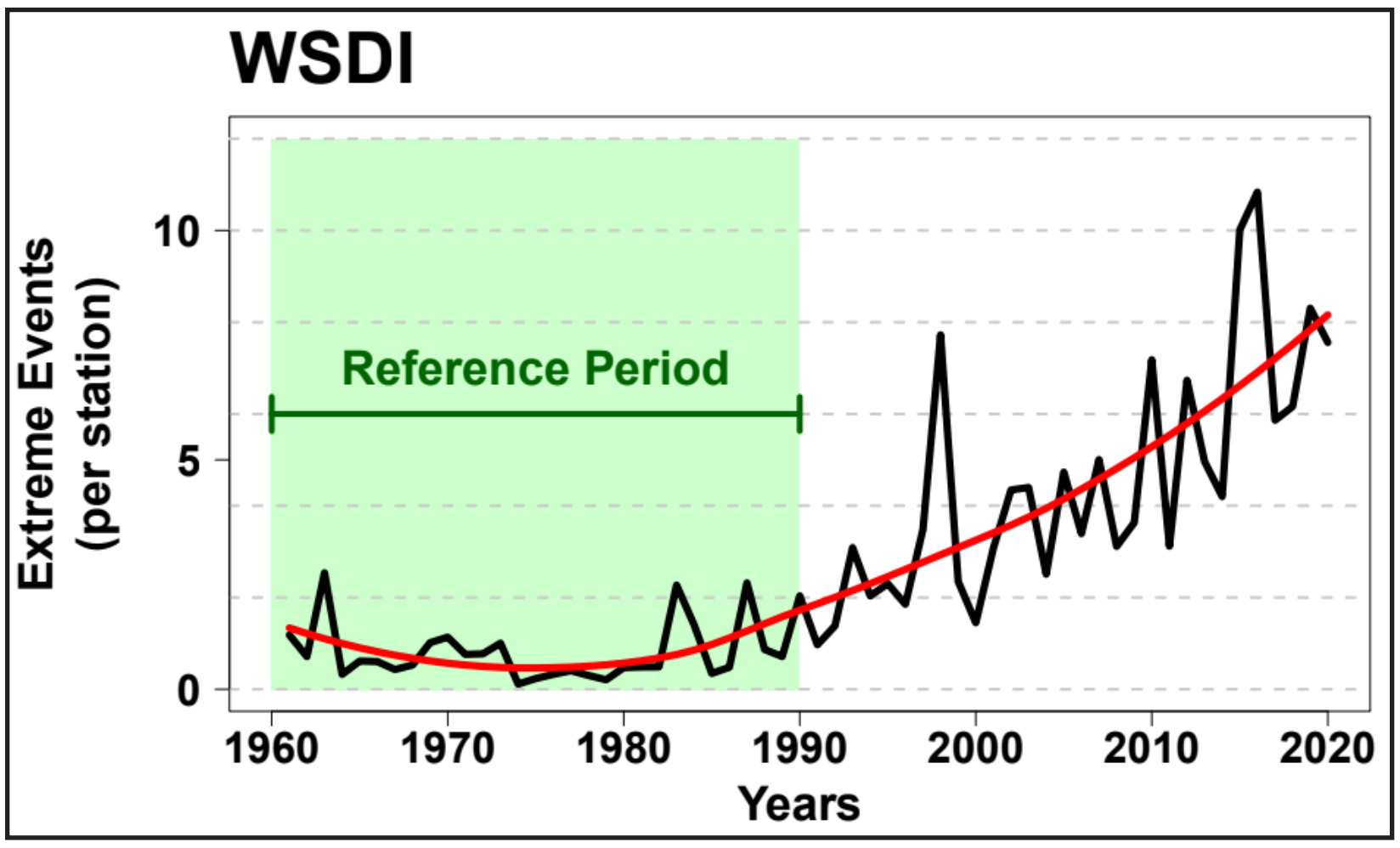

Figure 14 – Extreme TN90p (warm nights) events by represented as a time series (left) and over all stations (right). The time series shows a clear increase in events in Brazil, which is then confirmed by the positive trends in the majority of meteorological stations in the map

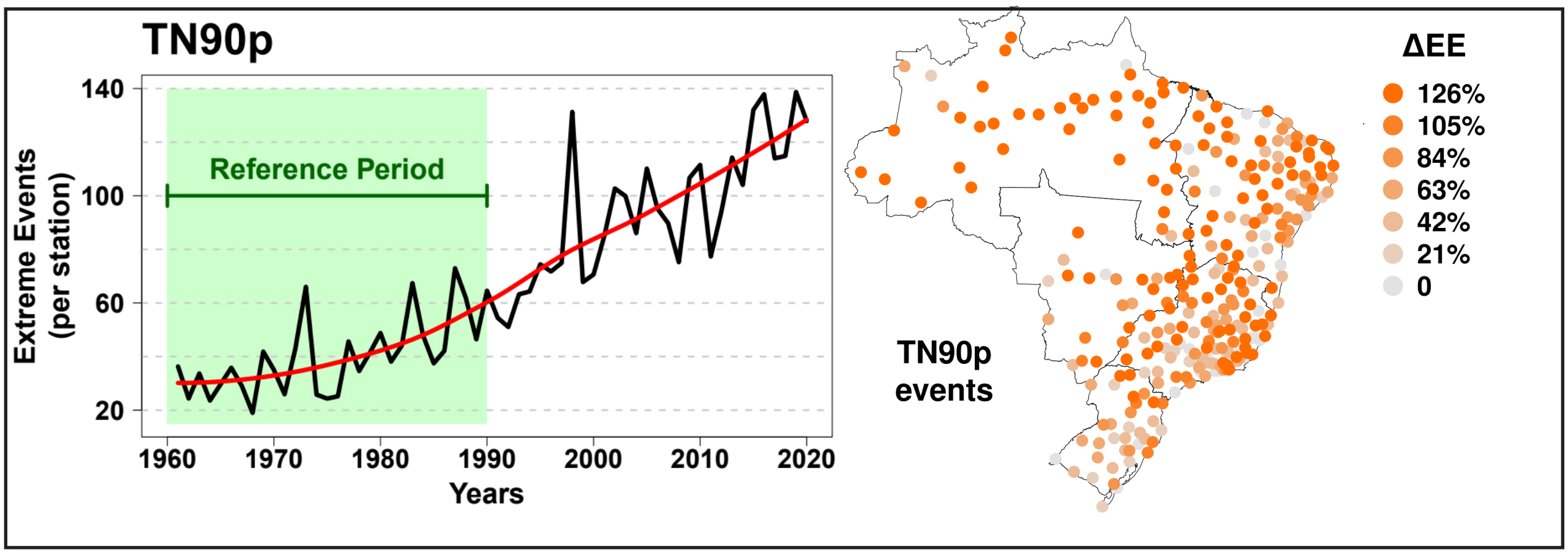

Figure 15 – Extreme TX10p (cold days) events by represented as a time series (left) and over all stations (right). The time series shows a clear decrease in events in Brazil, which is then confirmed by the negative trends in the majority of meteorological stations in the map

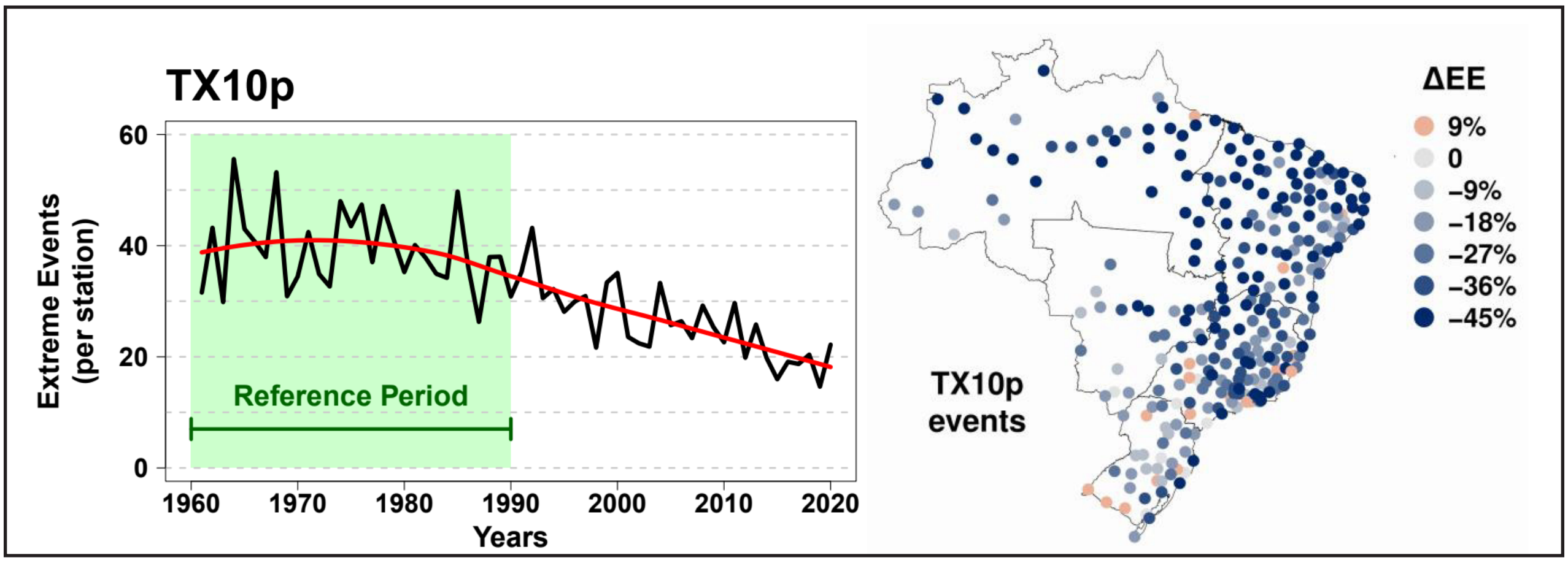

## Precipitation

Figure 16 shows the geolocalized precipitation trends for 6 indices. It is possible to see that for a bigger time window (12 months, SPI12 and SPEI12) the points are sparser, since gaps on the data make it difficult to compute the trends. Nevertheless, those trends confirm the overall behavior shown in Fig. 3 in the main text.

Figures 17, 18, 19 and 20 show the same geolocalized representation for additional indices. Figure 17 shows the map for the mRAI and RAI indices, with the colors representing variations in the Sen's slope. Figure 18 presents the map for an absolute index, which analyses the extreme precipitation events that present PR > 100mm and PR > 50mm (R50mm and R100mm). Figures 19 and 20 show the PRCPTOT and SDII indices, respectively, which take into account the yearly precipitations in each station. In all figures, the behavior seen is the same as the extreme precipitation map shown on the main text. Some, such as Fig. 18, show a little more noise since some meteorological stations do not have enough data (or precipitation) to present signal on this index. This, along with the SPI and SPEI indices, show the robustness of the data presented in the main text, where we find that the north and south regions seem to be getting wetter while the other regions seem to get drier, particularly northeast region, for all the indices.

Figure 16 – Trends on Standard Precipitation [Evapotranspiration] Indices (SP[E]I): Tendencies of drier/wetter precipitation series geolocalized in Brazil. The number behind the index indicates the time window considered to calculate the precipitation self-correlation series. The points indicate in green the meteorological stations that are becoming wetter, while purple points indicates the stations that are becoming drier

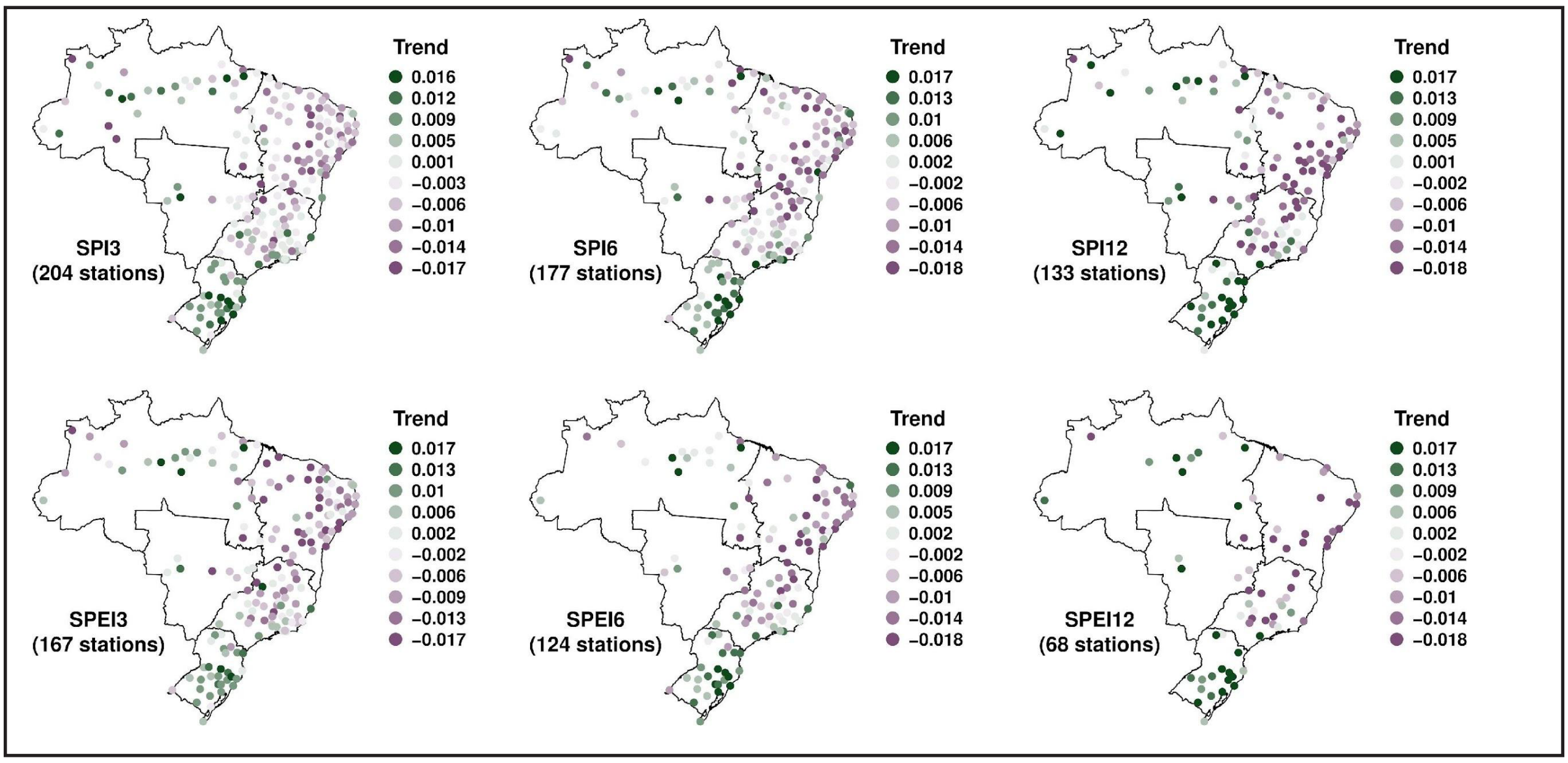

Figure 17 – Geolocalized trends of increasing (orange) and decreasing (darkblue) Sen's slope for the mRAI and RAI indices for each meteorological station. In both maps, there seems to be an increase in the slopes of stations in the North and South regions and a decrease in the Northeast and Southeast regions

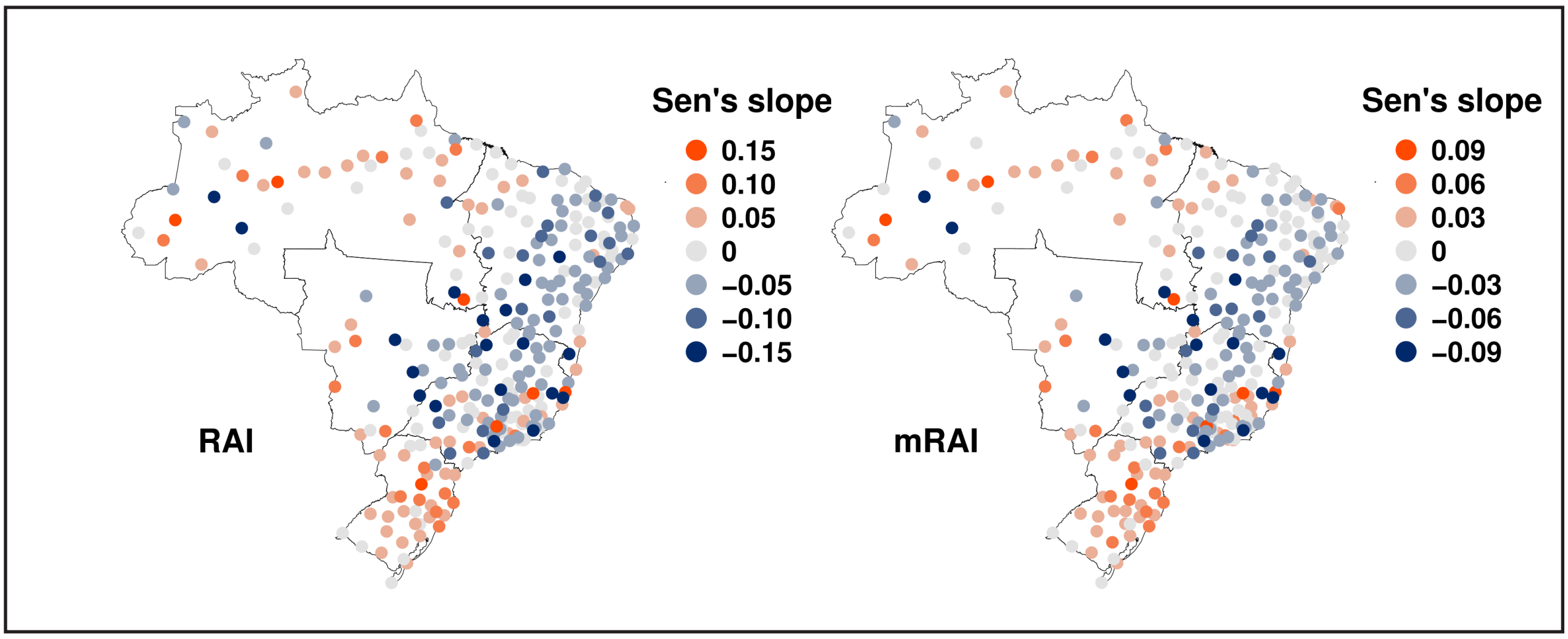

Figures 21 and 22 represent EE time series, measured as defined in Eq. 2, separated by different political regions in Brazil. The red lines represent each region's overall trend in events. As they are an average over multiple stations in each region, the behaviors shown are not as clear. Due to the average taken over a large region, trend behaviors may be swept away.

Figure 18 – Geolocalized trends of increasing (orange) and decreasing (darkblue) extreme events of precipitation above 100mm, on the left, and 50mm, on the right for each meteorological station. In both maps, there seems to be an increase in events in the North and South regions and a decrease in the Northeast and Southeast regions

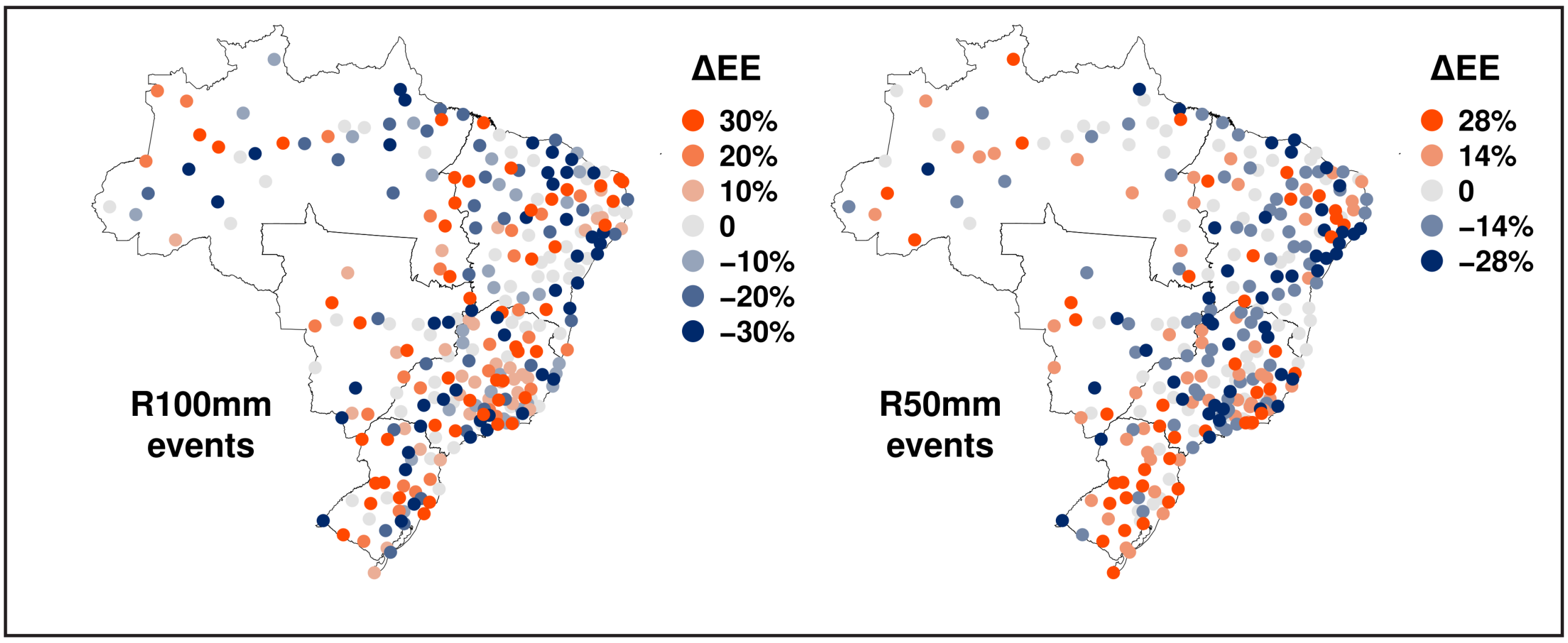

Figure 19 – Geolocalized trends of increasing (orange) and decreasing (darkblue) PRCPTOT events for each meteorological station. The overall trend seems to be of increasing events in the North and South regions and decreasing events in the Northeast and Southeast regions

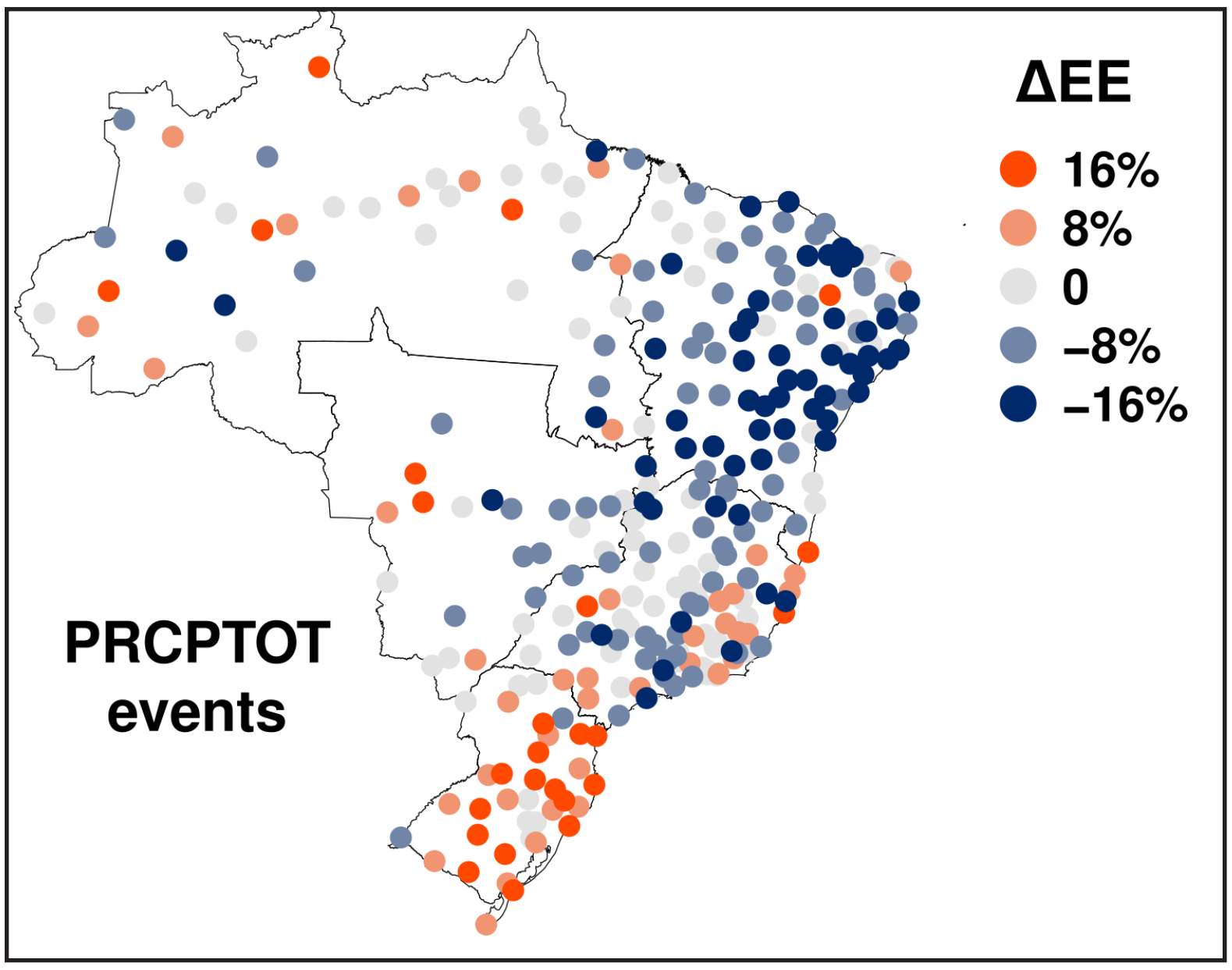

Figure 20 – Geolocalized trends of increasing (orange) and decreasing (darkblue) SDII events for each meteorological station. The overall trend seems to be of increasing events in the North and South regions and decreasing events in the Northeast and Southeast regions

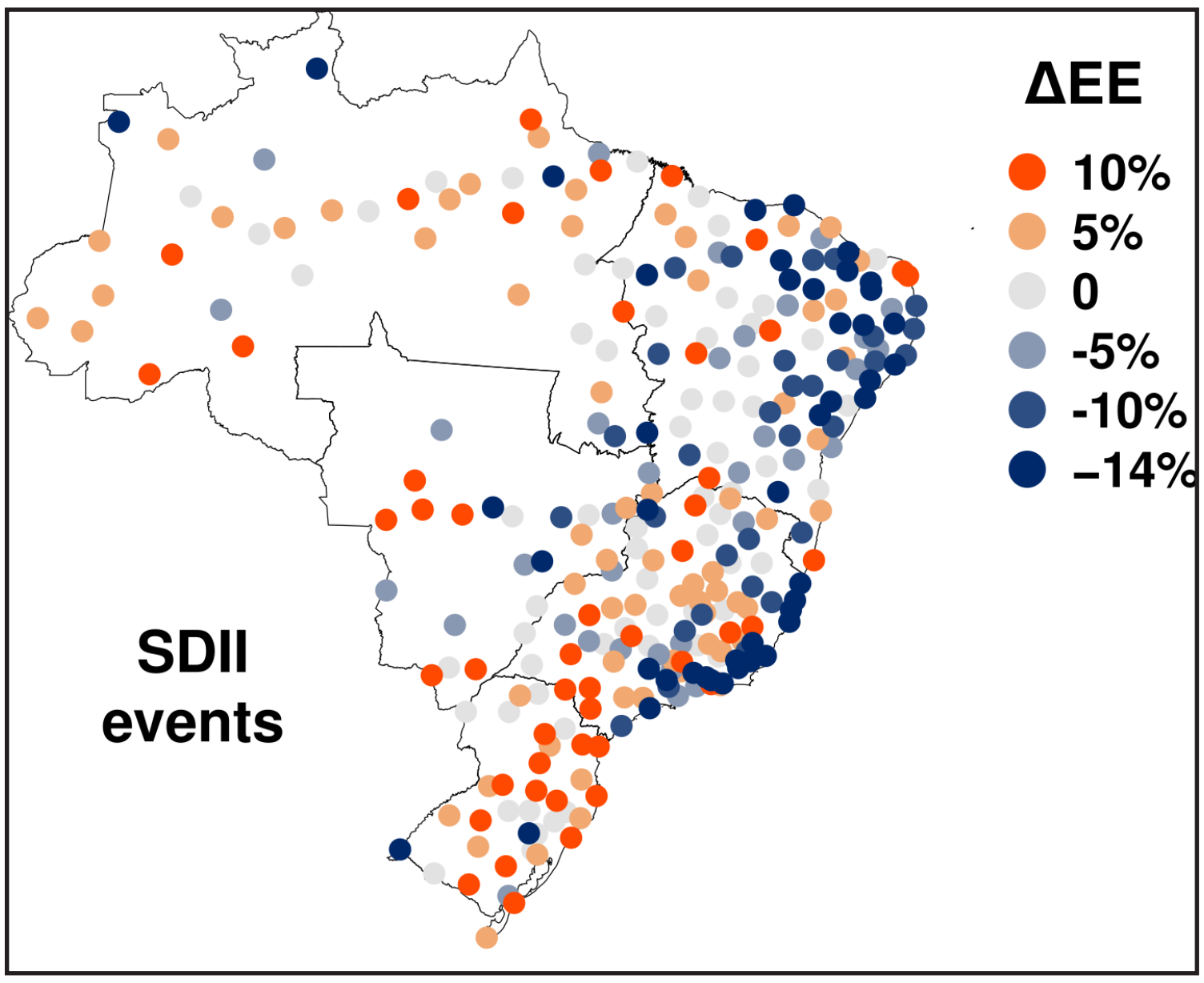

Figure 21 – Dry Days extreme events shown as time series separated by Brazilian political regions. The black dots indicate the yearly mean EE and the red lines indicate their trend

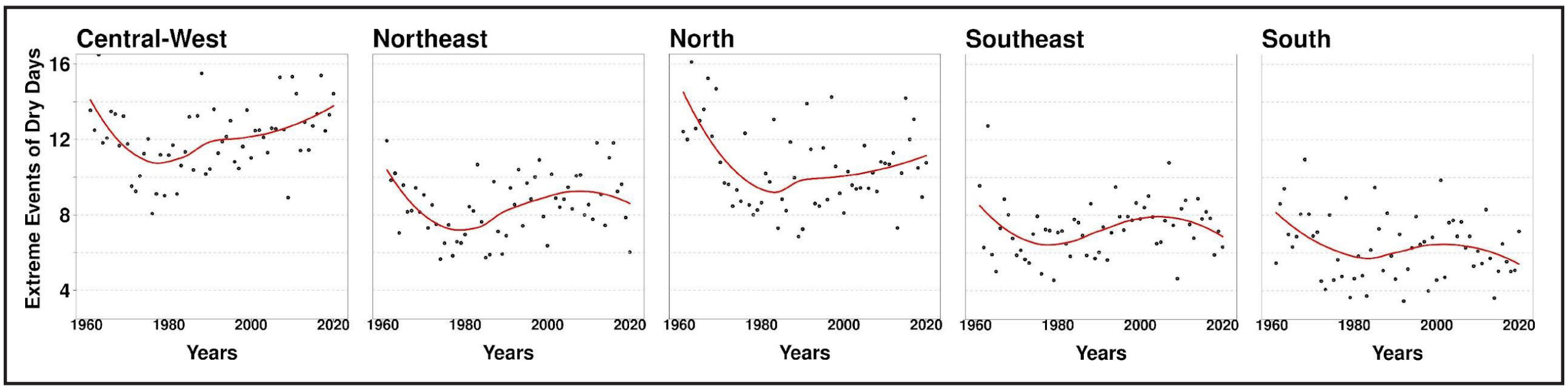

Figure 22 – Rain extreme events shown as time series separated by Brazilian political regions. The black dots indicate the yearly mean EE and the red lines indicate their trend

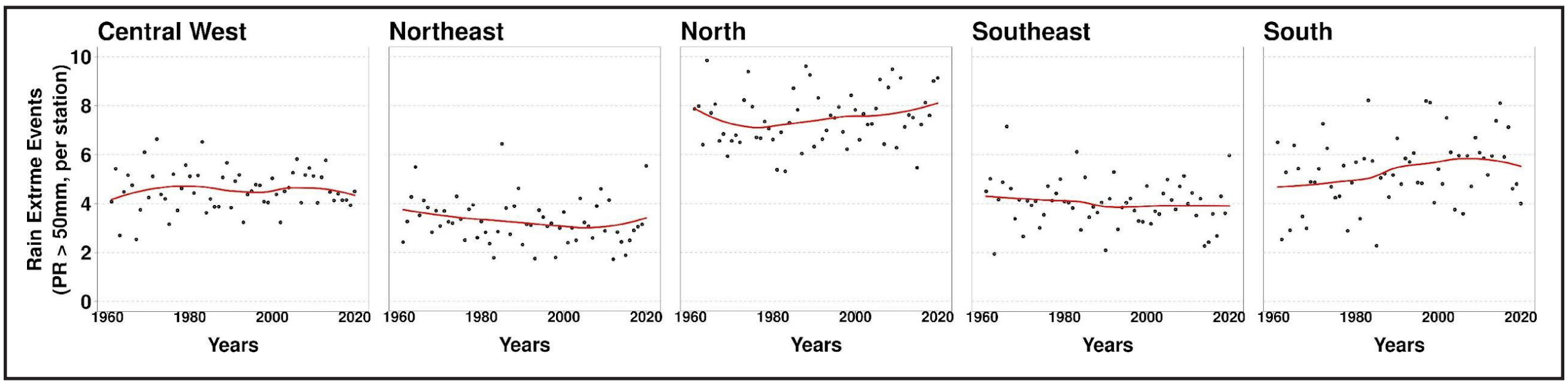

## REFERENCES

H.B. Mann. “Nonparametric tests against trend”. In: *Econometrica: Journal of the econometric society* (1945), pp. 245–259.

T.B. McKee et al. “The relationship of drought frequency and duration to time scales”. In: *Proceedings of the 8th Conference on Applied Climatology*. Vol. 17. 22. Boston, MA, USA. 1993, pp. 179–183.

P.K. Sen. “Estimates of the regression coefficient based on Kendall’s tau”. In: *Journal of the American statistical association* 63.324 (1968), pp. 1379–1389.

A. Tank, F. Zwiers, and X. Zhang. “Guidelines on Analysis of Extremes in a Changing Climate in Support of Informed Decisions for Adaptation”. In: *World Meteorological Organization* (Jan. 2009).

Charles Warren Thornthwaite. “An approach toward a rational classification of climate”. In: *Geographical review* 38.1 (1948), pp. 55–94.

S.M. Vicente-Serrano, S. Beguer´ıa, and J.I. Lo´pez-Moreno. “A multiscalar drought index sensitive to global warming: the standardized precipitation evapotranspiration index”. In: *Journal of climate* 23.7 (2010), pp. 1696–1718.

X. Zhang et al. “Avoiding inhomogeneity in percentile-based indices of temperature extremes”. In: *Journal of Climate* 18.11 (2005), pp. 1641–1651.

## Authorship contributions

### 1 – Davi Lazzari

Bachelor's degree in Physics from the Federal University of Rio Grande do Sul
https://orcid.org/0009-0006-5735-7643 • davi.lazzari@ufrgs.br
Contribution: Data collection, curation and analysis; Conceptualization and project administration; Methodology; Writing, reviewing and editing

### 2 – Amália Buchweitz Garcez

Bachelor's degree in Physics from the Federal University of Rio Grande do Sul
https://orcid.org/0009-0004-0959-6338 • amaliagarcez@gmail.com
Contribution: Data collection, curation and analysis; Methodology; Writing, reviewing and editing

### 3 – Nicole Magalhães Poltozi

Master's degree in Education from the University of Vale do Rio dos Sinos
https://orcid.org/0000-0002-9121-7592 • nicolemagalhaes3@gmail.com
Contribution: Data collection, curation and analysis

### 4 – Gianluca de Souza Pozzi

Bachelor's degree in Geography from the Federal University of Rio Grande do Sul
https://orcid.org/0000-0002-1298-839X • gianlucapozzi1999@gmail.com
Contribution: Data collection, curation and analysis

### 5 – Carolina Brito

PhD in Physics from the Federal University of Rio Grande do Sul
https://orcid.org/0000-0003-1033-7590 • carolina.brito@ufrgs.br
Contribution: Data collection, curation and analysis; Conceptualization and project administration; Methodology; Writing, reviewing and editing; Project supervision

## How to quote this article